\documentclass[aps,pra,preprint,groupedaddress]{revtex4}

\usepackage{graphicx}

\usepackage{amsmath,amsthm,amsfonts, latexsym}

\newcommand{\mtx}[2]{\left(\begin{array}{#1}#2\end{array}\right)}

\def\lsim{\mathrel{\rlap{\lower4pt\hbox{\hskip1pt$\sim$}}
    \raise1pt\hbox{$<$}}}

\begin{document}


\title{Real-Vector-Space Quantum Theory \\ with a Universal Quantum Bit}


\author{Antoniya Aleksandrova$^{1,2}$, Victoria Borish$^{1,3}$, and William K.~Wootters$^{1}$}
\affiliation{$^{1}$Department of Physics, Williams College, Williamstown, MA 01267, USA \\
$^{2}$Theory of Condensed Matter Group, Cavendish Laboratory, University of Cambridge, Cambridge CB3 0HE, UK \\
$^{3}$Institute for Quantum Optics and Quantum Information, Austrian Academy of Science, Boltzmanngasse 3, A-1090 Vienna, Austria}


\begin{abstract}
We explore a model of the world based on real-vector-space quantum theory.  In our model the familiar complex phase
appearing in quantum states is replaced by a single binary object that we call the ubit, which is not localized and which can interact with any object in the world.  Ordinary complex-vector-space quantum theory can be recovered from this model if we simply
impose a certain restriction on the sets of allowed measurements and transformations (Stueckelberg's rule), but in this paper we
try to obtain the standard theory, or a close approximation to it, {\em without} invoking such a restriction.  
We look particularly at the effective theory that applies to a subsystem when the ubit is 
interacting with a much larger environment.  In a certain limit it turns out that the ubit-environment interaction
has the effect of enforcing Stueckelberg's rule automatically, and we obtain a one-parameter family of
effective theories---modifications of standard quantum theory---that all satisfy this rule.  
The one parameter is the ratio $s/\omega$, where $s$ quantifies the strength of the ubit's interaction with the
rest of the world and $\omega$ is the ubit's rotation rate.  We find that when this parameter is small but not zero, the
effective theory is similar to standard quantum theory but is characterized by spontaneous decoherence of isolated systems.   

\end{abstract}


\maketitle


\section {Introduction: Real and Complex Quantum Theory}

Standard quantum theory is based on a complex Hilbert space: density matrices, observables and
reversible transformations are all represented by linear operators on such a space.  However, it has been known since the early days of quantum mechanics
that many features of the theory are shared by two alternative, hypothetical theories, in which the complex Hilbert space is replaced by either a 
real or a quaternionic Hilbert space.  For example, in their analysis of the logical structure of quantum theory in 1936, Birkhoff and von Neumann noted explicitly
that their postulates, which were intended to capture this logical structure, would be satisfied just as well by the real and quaternionic models as by the 
standard complex theory \cite{Birkhoff}.  To be sure, from an empirical point of view there has hardly been any contest among these three theories: the complex version 
has survived every test, and no experiment has been done that requires either the real theory or the quaternionic theory for its explanation.  Nevertheless, over the years researchers have sought a more fundamental understanding---not merely an empirical understanding---of
the origin of the complex structure \cite{Bohm, Stueckelberg1, Stueckelberg2, Mackey, Trautman, Lahti, Gibbons, Gibbons2, Barbour, Hardy, Caves, Aaronson, Goyal, DAriano}.  

One avenue of investigation along these lines was carried out around 1960 by Stueckelberg \cite{Stueckelberg1, Stueckelberg2}.  He began with the representation of probabilities as squares of real amplitudes, treating this step as a natural generalization of ordinary probability theory.  He felt that one needed to 
explain the complex structure somehow, starting from a real Hilbert space.  In order to provide such an explanation he imposed the requirement that the theory admit an uncertainty principle of a specific form. 
This requirement led him to introduce a special operator $J$ to be used in the expression of the uncertainty principle.  The operator $J$
has the property that its square is the negative of the identity operator \footnote{A linear operator with this property is said to 
define a {\em complex structure} on a real vector space \cite{Kobayashi, Trautman}.}, and the uncertainty principle then holds if one requires every observable to commute with $J$.
With this restriction on the observables, the theory becomes equivalent to standard complex
quantum theory, with Stueckelberg's special operator playing the role of the complex number $i$.  We will say that a real operator 
satisfies ``Stueckelberg's rule'' if it commutes with $J$. 

In the present paper we take the real-vector-space theory seriously as a potential theory of nature and, like Stueckelberg, we consider the possibility that the complex structure is somehow to be located or embedded in the real theory \footnote{One appealing feature of the 
real-vector-space theory, at least in the finite-dimensional case, is this:  the transfer of information from the preparation of a pure state to the subsequent outcome
of a complete orthogonal measurement is {\em optimal}---in a specific sense---relative to other conceivable probabilistic laws relating the preparation to the outcomes \cite{Wootters2}.  One does not see this optimization in the standard complex theory.}.  However, we do not want simply to impose Stueckelberg's rule.  Rather, we ask whether the complex structure might emerge dynamically in a particular model.  To get started we recall how one can express in real-vector-space terms the basic structure of quantum theory with a finite-dimensional state space.
The finite-dimensional case requires a step that is not needed in 
the infinite-dimensional case: we have to double the Hilbert-space dimension.  That is, to model an ordinary quantum system with 
a $d$-dimensional state space, we need a 
real Hilbert space of 
$2d$ dimensions.

Suppose, for example, that one wants to describe only the spin of a spin-1/2 particle.  A real state vector in a two-dimensional state space is clearly inadequate.  To get all the allowed quantum states, one needs to double the dimension to four.  But a four-dimensional real vector space is not the same as a two-dimensional complex
space: one needs three real numbers to specify a rank-one projection operator (a pure state) in the real space but only two real numbers to specify such an operator
in the complex space (these could be the two angular coordinates of the Bloch sphere).  In order to restrict the set of states, one can impose a version of Stueckelberg's rule 
\footnote{In Refs.~\cite{Stueckelberg1} and \cite{Stueckelberg2}, Stueckelberg did not explicitly restrict the set of states but only the observables and transformations.  However, his prescription for expressing a complex inner product in terms of real inner products (Ref.~\cite{Stueckelberg2}, p.~747) generates the same probabilities as would be produced by a mixed state that commutes with $J_U \otimes I_A$.}, namely, that all density matrices commute with the matrix
\begin{equation}  \label{oneone}
J \otimes I =\mtx{cc}{0 & -1 \\ 1 & 0}  \otimes \mtx{cc}{1 & 0 \\ 0 & 1}  = \mtx{cccc}{0 & 0 & -1 & 0 \\ 0 & 0 & 0 & -1 \\ 1 & 0 & 0 & 0 \\  0 & 1 & 0 & 0},
\end{equation}
which has the property that $(J \otimes I)^2 = - I \otimes I $.  (We find it convenient to use the symbol $J$ for the $2 \times 2$ matrix rather than
for the larger matrix as in Stueckelberg's papers or in the mathematical literature on complex structures.)  As we will see in Section II, the theory that results from applying this restriction not only to states but also to measurements and transformations is equivalent to standard quantum theory for this system, and a similar result holds for any other system, regardless of the dimension of the state space.  One always needs to double the dimension in order for the real-vector-space theory to be able to accommodate all the states and operations of the complex theory, and then one needs to impose a restriction so as to limit the sets of states and operations in the right way.

Now, doubling the dimension corresponds to adding to the system a single binary quantum object.  So we can say that the spin of a spin-1/2 particle, that is, a qubit, is equivalent to a binary real-vector-space object---a rebit---together with an auxiliary rebit, such that the whole system obeys Stueckelberg's rule.  One might initially think, then, that to describe the spins of $n$ spin-1/2 particles, one would need, in addition to the $n$ basic rebits, another $n$ auxiliary rebits to turn them all into qubits.  But it is clear that this is not the case; a single auxiliary rebit is sufficient for the whole system, because it is all that is needed to double the dimension.  This fact has been noted by a number of 
authors \cite{Rudolph, Fernandez, McKague1, McKague2}.  For example, it has been shown how one could simulate an $n$-qubit quantum computation by a real-vector-space quantum computation involving only $n+1$ rebits \cite{Rudolph, Fernandez}.  

We are thus led to consider the following model.  Every system is to be described as a quantum object in a real vector space, with the same dimension it would normally have in the complex theory, and in addition, there is a single auxiliary rebit.  We call this auxiliary rebit the universal rebit, or ubit, because in this model it needs to be able to interact with every object in the world.  By invoking Stueckelberg's rule, we could, not surprisingly, make our model equivalent to standard quantum theory, as we will show in Section II 
\footnote{To get {\em quaternionic} quantum
theory starting from the real theory, one could add {\em two} ubits and then impose the rule that all 
operators---including density matrices and the operators representing measurement outcomes---commute
with
the pair of two-ubit operators $I \otimes J$ and $J \otimes X$, where $X$ is the usual Pauli matrix.  Addition and multiplication of the resulting $4 \times 4$ matrices
on the two-ubit space are then equivalent to the addition and multiplication of quaternions.  
A thorough development of quaternionic quantum
theory can be found in Ref.~\cite{Adler}.}.
However, as we have said, in this paper we want to take a different tack: we ask
whether we can arrive at ordinary quantum theory, or an approximation to ordinary quantum theory, {\em without} invoking Stueckelberg's rule.  If the ubit is interacting with everything, then no local observer will be able to control its interactions with distant objects.  It is conceivable that this uncontrollability could lead to an {\em effective} theory that approximates ordinary quantum theory, even though the underlying theory
is the real-vector-space theory.  Whether such an approximation is possible is the question with which we begin our investigation.

One finds that with no further assumptions, a random interaction between the ubit and a large environment does not reproduce standard quantum theory.  Rather, the ubit quickly factors out of the system and becomes irrelevant, and one is left with ordinary real-vector-space quantum theory with no ubit, a theory that is in serious conflict with experiment.  However, the results we present in this paper indicate that if the ubit is {\em rotating} sufficiently rapidly in its two-dimensional real vector space (rotation is the only internal dynamics
possible for this simple system), then one does recover an effective theory 
that is very much like ordinary quantum theory.
Our main goal in this paper is to begin to discern the features of this effective theory.  One feature we might expect to see, and we do indeed see, is that an isolated system can undergo spontaneous decoherence with an associated increase in entropy, even when no decoherence would be predicted by standard quantum theory.  One expects such decoherence because, although the system may be isolated in the sense that it does not experience any ordinary interactions, the ubit is never isolated and can therefore serve as a conduit of information to the rest of the world. 

It may seem quite fanciful to imagine a special rebit with no particular location, associated with the universe as a whole.  Indeed in this paper we are not prepared to offer any interpretation of this object beyond what the mathematics implies.  But we note that ordinary quantum theory does have a feature that is something like the ubit.  For a quantum system with definite energy $E$, even if it is a spatially extended system with many parts, the time dependence of the Schr\"odinger wavefunction is expressed in an overall factor $e^{-iEt/\hbar}$ multiplying the rest of the wavefunction.  It is interesting that there is only one such time-dependent phase factor for the whole system, not one for each part.  (Each part may not have a definite energy of its own.)  Moreover the phase factor does represent a rotation in a certain two-dimensional real vector space, namely, the complex plane.  In our real-vector-space model there is no phase factor, but in its place there is the ubit, with its own 
internal dynamics (that is, the rotation) and its own interactions with other systems.  

One might worry that our model is immediately suspect in that it seems to allow instantaneous communication over an arbitrary distance: a sender Alice could allow her particle $A$ to interact with the ubit, which immediately interacts with particle $B$ in a distant galaxy, delivering Alice's message to Bob (who has somehow managed to be there).  In this paper, though, we focus particularly on a limiting case in which 
it seems that such instantaneous communication cannot occur.  There are three relevant frequency scales in the model: (i) the rotation rate of the ubit, (ii) the typical strength of interaction between the ubit and the large environment, and (iii) the typical frequency scale of the local dynamics.  We focus our attention on the case in which the first two of these frequency scales approach infinity with a fixed ratio, while the third remains finite.  In this limit we use a heuristic argument to obtain a reduced dynamics of the local system.  Under our reduced dynamics we automatically recover Stueckelberg's rule for the states and transformations, which we show prevents
instantaneous signaling through the ubit.  
The ratio between the strength $s$ of the ubit's interaction with the environment and the ubit's rotation rate 
$\omega$ serves as a single parameter that characterizes the effective theory.  When that ratio is zero, our results 
indicate that one recovers ordinary quantum theory.  Our 
primary interest in this paper is the case when the ratio is small but not zero.   

Though our model does not allow instantaneous communication under the conditions we consider, the model itself is manifestly nonlocal
and quite contrary to the spirit of special relativity (not to mention general relativity).  At any given value of the universal
time coordinate, the 
ubit undergoes the same change everywhere.
One can imagine making the interactions local 
by replacing the ubit with a ``ubit field.''  It is interesting to ask
whether such a change would ruin the partial agreement we will find with standard quantum theory.  As we will see, it is important
in our model that the ubit be interacting with a large environment.   Over any short time interval, a {\em ubit field} at a given location 
would interact with an
environment of limited extent, so the ``beneficial'' environmental effects that we rely on would also be limited.  We do not explore this question in the present
paper but confine our attention to the simple model with a single, binary ubit.  
Despite the underlying nonlocality, the fact that we can get 
an effective theory displaying only local interactions
makes it seem worth exploring 
the model to see where it leads.  

The paper is organized as follows.  In Section II we specify what we mean by ``real-vector-space quantum theory,'' and we demonstrate the equivalence with ordinary quantum theory when Stueckelberg's rule is imposed with no further restrictions.  (We do not impose Stueckelberg's rule in the later sections.)  In Section III we investigate, numerically and analytically, the dynamics of the ubit interacting randomly with an environment.  We find, among other things, that any component of the ubit state that fails to commute with $J$ quickly decays.  Our next step is to restrict our attention to the limiting case described above in which this decay happens instantaneously and continually.  In Section IV we consider one specific physical example, the spin of a spin-1/2 particle precessing in a magnetic field, and we explore its behavior numerically when 
the problem is recast in the ubit model.  We identify three ways in which this behavior deviates from standard quantum theory: (i) the frequency of precession is reduced; (ii) there is a long-term dephasing (mentioned above); (iii) there is a periodic variation in 
the {\em purity} of the spin state, indicating that the spin is periodically becoming correlated and then uncorrelated with the environment.  In Section V
we explain all these effects analytically using perturbation theory, taking the ratio ${s}/\omega$ as our small
parameter.  We find, though, that at least to second order in this parameter we can eliminate the strange oscillation in purity
simply
by reinterpreting the theory.  The reinterpretation---which does not eliminate either the reduction in precession frequency
or the decoherence---is presented in Section VI.  We consider
in that section systems with higher state-space dimension than a spin-1/2 particle, but in this paper we do not analyze the higher-dimensional case in detail.  Section VII focuses on the fact that in the ubit model
there is not a unique mapping from the complex theory to the real theory. The choice of mapping amounts to an additional specification of the dynamics beyond what is determined by the Hamiltonian.  It turns out that one particular choice would
render the retardation in the evolution unobservable---it would slow all processes by the same factor---leaving only the decoherence as 
a potentially observable effect of the ubit model.  We discuss our results and draw conclusions in Section VIII.     

\section{Real-Vector-Space Quantum Theory with Finite Dimension}

One can identify four main components of the basic framework of standard quantum theory in a complex vector space: (i) States are represented by positive semi-definite operators with unit trace (density matrices). (ii) A reversible evolution is represented by a unitary transformation. (iii) An ideal repeatable measurement is represented by a set of orthogonal  projection operators $\Pi_i$, such that 
the probability of the outcome $i$ when the state is $\rho$ is $\hbox{Tr}(\Pi_i\rho)$, and when outcome $i$ occurs, the final state of the system is proportional to $\Pi_i\rho \Pi_i$. (iv) The state space of a composite system is the tensor product of the state spaces of the components.  Other kinds of evolution and measurement are certainly possible, but they can be derived from the above kind by applying these rules to a larger system and then considering the effects on a subsystem.  

The analogous statements for real-vector-space quantum theory are exactly the same, except that all the operators are real.  In particular this means that a reversible evolution is represented by an {\em orthogonal} transformation, which is the real version of a unitary 
transformation.

Let us now write down the differential equation governing the evolution of a state in real-vector-space quantum theory.  In the
usual complex theory, we can write the equation of evolution as 
\begin{equation}
\frac{d\rho}{dt} = [-iH/\hbar, \rho].
\end{equation}
In the real-vector-space theory there is
no direct analogue of the Hamiltonian $H$, but we can replace the antihermitian operator $-iH/\hbar$ with an {\em antisymmetric} real operator $S$, so that
the evolution equation becomes
\begin{equation}  \label{diffeq}
\frac{d\rho}{dt} = [S, \rho].
\end{equation}
We will call the operator $S$ the ``Stueckelbergian'' of the system.  If the Stueckelbergian is constant, as we will always assume in
this paper, then the general solution of Eq.~(\ref{diffeq}) is
\begin{equation}
\rho(t) = e^{St} \rho(0) e^{-St}.
\end{equation}
As $S$ is antisymmetric, the operator $e^{St}$ is orthogonal.

We now show how one can recover standard quantum theory from the real-vector-space version by adding the ubit and imposing Stueckelberg's rule.  Much of what follows in this section 
(minus the interpretation in terms of a ubit) is similar to the account given in Ref.~\cite{Myrheim}.  

Suppose the system we want to describe has a $d$-dimensional (complex) Hilbert space.  Then we start by considering a $d$-dimensional real-vector-space object $A$ together with the ubit $U$.  Consider any matrix $M$ that might apply to the $UA$ system, whether it be a density matrix, an orthogonal evolution operator or antisymmetric Stueckelbergian, or the projection operator associated with a measurement outcome.  The $2d \times 2d$ matrix $M$ can be written as 
\begin{equation}
M = \mtx{cc}{M_{00} & M_{01} \\ M_{10} & M_{11}},
\end{equation}
where each $M_{jk}$ is a real $d \times d$ matrix and the subscripts ``0'' and ``1'' refer to a pair of 
orthogonal basis vectors in the ubit's two-dimensional space.  
We now impose Stueckelberg's rule: we insist that $M$ commute with $J_U \otimes I_A$, where $J_U$ is the $2 \times 2$ matrix we mentioned in the introduction and $I_A$ is the $d \times d$ identity.  (We include the alphabetic subscripts to indicate which system the operator acts on.)  
That is, we insist that
\begin{equation}
\mtx{cc}{M_{00} & M_{01} \\ M_{10} & M_{11}}\mtx{cc}{0 & -I_A \\ I_A & 0} = \mtx{cc}{0 & -I_A \\ I_A & 0}\mtx{cc}{M_{00} & M_{01} \\ M_{10} & M_{11}},
\end{equation}
which implies that $M_{00} = M_{11}$ and $M_{10} = - M_{01}$.  Thus we can write 
\begin{equation}
M = \mtx{cc}{M_{00} & -M_{10} \\ M_{10} & M_{00}}.
\end{equation}

We can map any such matrix into a smaller, $d \times d$ complex matrix, such that under this mapping, the laws of 
real-vector-space quantum theory become the laws of complex-vector-space quantum theory.  The mapping is this: for
a matrix $M$ representing an orthogonal transformation, a Stueckelbergian, or a projection operator, we have
\begin{equation}  \label{mapping1}
M=\mtx{cc}{M_{00} & -M_{10} \\ M_{10} & M_{00}}\hspace{3mm} \rightarrow \hspace{3mm}{\mathcal M} = M_{00} + iM_{10},
\end{equation}
and for a density matrix $\rho$, we have
\begin{equation}  \label{mapping2}
\rho = \mtx{cc}{\rho_{00} & -\rho_{10} \\ \rho_{10} & \rho_{00}}\hspace{3mm} \rightarrow\hspace{3mm} {\mathcal \sigma} = 2(\rho_{00} + i\rho_{10}).
\end{equation}
The special rule for density matrices is simply to make sure every real or complex density matrix has unit trace.  Note that $\rho_{10}$ must be an antisymmetric matrix in order that $\rho$ be symmetric.  This implies that $\hbox{Tr}\,\rho = \hbox{Tr}\,\sigma$,
since $\rho_{10}$ is traceless.

One can verify that Eq.~(\ref{mapping1}) faithfully preserves matrix multiplication: if $M_1 \rightarrow {\mathcal M}_1$ and $M_2 \rightarrow {\mathcal M}_2$, then $M_1 M_2 \rightarrow {\mathcal M_1}{\mathcal M_2}$.  
Moreover Eqs.~(\ref{mapping1}) and (\ref{mapping2}) together preserve the trace of a density matrix times a projection operator:
if $\rho \rightarrow \sigma$ and $\Pi \rightarrow \Upsilon$, then $\hbox{Tr}\,\Pi\rho = \hbox{Tr}\,\Upsilon \sigma$.
These two facts guarantee that the real-vector-space laws (restricted by Stueckelberg's rule) are equivalent to the complex-vector-space laws under this mapping.  

Eqs.~(\ref{mapping1}) and (\ref{mapping2}) show how to convert real matrices that satisfy Stueckelberg's rule into
complex matrices.  
One can just as easily go the other way around.
For example, given a complex Hamiltonian $H$, we can write the corresponding Stueckelbergian $S$ as
\begin{equation}  \label{HtoS}
S = \mtx{cc}{\hbox{Re}(-iH/\hbar) & - \hbox{Im}(-iH/\hbar) \\ \hbox{Im}(-iH/\hbar) & \hbox{Re}(-iH/\hbar)}  = I_U \otimes \hbox{Re}(-iH/\hbar) + J_U \otimes \hbox{Im}(-iH/\hbar).
\end{equation}
Given a complex density matrix $\sigma$, one obtains the corresponding real density matrix $\rho$ by performing a similar 
operation \footnote{Alternatively, we could write 
$\rho = (1/2)[(1/2)(I_U -Y_U) \otimes \sigma + (1/2)(I_U +Y_U) \otimes \bar{\sigma}]$, where $Y$ is the imaginary Pauli matrix
and the bar indicates complex conjugation.  This
form is an extension of the general prescription given in Ref.~\cite{BRS} for bringing a reference frame into the quantum description.
In this case the ubit would be the reference frame.  We do not use this formulation here because we want to emphasize
the possibility of treating the ubit as a rebit rather than as a qubit.  For a rebit, we cannot have $(I\pm Y)/2$ as possible states
along with the states $(I\pm X)/2$ and $(I\pm Z)/2$, which we will need.}:
\begin{equation}  \label{complextoreal}
\rho = \frac{1}{2}\mtx{cc}{\hbox{Re}\, \sigma & - \hbox{Im}\, \sigma \\ \hbox{Im}\, \sigma & \hbox{Re}\, \sigma} = \frac{1}{2}\left(
I_U \otimes \hbox{Re}\, \sigma + J_U \otimes \hbox{Im}\, \sigma \right).
\end{equation}
Note that in this real-vector-space setting, under 
the restriction that $\rho$ commute with $J_U \otimes I_A$, no system is described by a state vector.  In fact the
purity $\hbox{Tr}\,\rho^2$ cannot be greater than $1/2$: from Eq.~(\ref{complextoreal}), we have $\hbox{Tr}\,\rho^2 
= (1/2) \hbox{Tr}[ (\hbox{Re}\,\sigma)^2 - (\hbox{Im}\,\sigma)^2 ] =  (1/2) \hbox{Tr}\,\sigma^2 \le 1/2$.  Thus every state has to 
be represented by a density matrix, even if it corresponds to a pure state in standard quantum theory, and the minimum rank of 
any density matrix is two. 

Of course 
the requirement that operators commute with $J_U \otimes I_A$
is crucial here.  Without this restriction, one could indeed have a state vector in the real-vector-space theory---in a sense such a state would be purer than any pure state in standard quantum theory.  Also note that a general orthogonal matrix in $2d$ dimensions can be characterized by $(2d^2 - d)$
real parameters, whereas a unitary matrix in $d$ dimensions requires only $d^2$ real parameters.  So hardly any of those
orthogonal matrices correspond to unitary matrices. This is one sense in which the unrestricted real-vector-space theory
allows too many possibilities.  
In the following sections we will not impose Stueckelberg's rule but will try to achieve its effects in another way. 

It is useful to note that, in the absence of any restrictions, every real matrix $M$ acting on the $UA$ system can be broken uniquely into two parts $M_c$ and $M_a$, which respectively commute
and anticommute with $J_U \otimes I_A$.  We can write the two parts as 
\begin{equation}
M_c = \frac{1}{2}\left[ M - (J_U \otimes I_A) M (J_U \otimes I_A) \right]
\end{equation}
and
\begin{equation}
M_a = \frac{1}{2}\left[ M + (J_U \otimes I_A) M (J_U \otimes I_A) \right].
\end{equation}
In these terms one can see, for example, that if all observables and transformations commute with $J_U \otimes I_A$, then the anticommuting
part of a density matrix, $\rho_a$, can have no observable effects.  Let an initial density matrix $\rho$ be transformed by an orthogonal
transformation $O$ and then tested for a property represented by the projection operator $\Pi$.  If $O$ and $\Pi$ commute with
$J_U \otimes I_A$, then the contribution from $\rho_a$ to the probability of the ``yes'' outcome is
\begin{equation}
\hbox{Tr}(\Pi O\rho_aO^T) = (1/2) \hbox{Tr}\left\{ \Pi O\left[\rho + (J_U \otimes I_A) \rho (J_U \otimes I_A) \right] O^T \right\}= 0.
\end{equation}
(In the second term inside the trace, we can move one factor of $J_U \otimes I_A$ through $O$, $\Pi$, and $O^T$ so that it 
combines with the other factor of $J_U \otimes I_A$ to yield $-I_U \otimes I_A$.  The two terms then cancel.)  
Thus if we impose Stueckelberg's rule
on all observables and transformations, the physical predictions of the theory will not depend on whether we also impose
this restriction on the set of allowed states.

\section{The ubit interacting with the environment}

\subsection{Specification of the model}

Ultimately we want to consider a system $A$ interacting with the ubit $U$, which is also interacting with an environment $E$
(but there will be no direct interaction between $A$ and the environment).  
We will take the Stueckelbergian of the entire system to be of the form
\begin{equation}  \label{wholeS}
\hat{S} = -\omega I_E \otimes J_U \otimes I_A + {s} B_{EU} \otimes I_A + I_E \otimes S_{UA},
\end{equation}
where the subscripts again indicate the system on which each operator acts and we use a hat (rather than the subscript
$EUA$) to distinguish
those operators that act on the entire system.  The operator $J_U$ generates rotations of the ubit, so $\omega$
is the ubit's rotation rate.  The operator $B_{EU}$ characterizes both the interaction of the ubit with the environment and the internal dynamics
of the environment itself, and ${s}$ determines the scale of these interactions.  Finally the operator $S_{UA}$ is the local Stueckelbergian, the one part of $\hat{S}$ that we imagine
can be controlled by an observer.  

This last part deserves some discussion.  The ubit is not localized, but we are assuming that it is available to be manipulated and measured
by any observer.  That is, our local observer---whom we will call Alice---can arrange for the implementation of an arbitrary 
Stueckelbergian $S_{UA}$ involving the ubit
and the local real-vector-space object $A$ and can measure any observable on the $UA$ system.  One might worry that different physical systems all over the universe will be competing
to achieve contradictory effects on the same ubit.  Indeed we will see that something along these lines does happen.  It will turn out that the 
interaction of the ubit with the environment severely limits what Alice will actually be able to do.  However, we do not impose any
such restriction in the basic model.  

We now describe our method of generating the matrix $B_{EU}$.  
One could reasonably model the environment as a collection of, say, rebits or higher-dimensional objects, each having some random interaction with the ubit 
but no interaction with each other.  However, when we do our numerical experiments, we would like each run to be reasonably reproducible, as it would be for a very large environment; so we want $B_{EU}$ to include as many randomly chosen parameters as possible without having to make the 
environment's dimension intractably large.  We therefore model the environment as a single system, with the matrix $B_{EU}$ simply chosen at random.
More precisely, taking $N$ to be the Hilbert space dimension of the environment, we 
create a $2N \times 2N$ matrix $R$, each component chosen uniformly between $-1$ and $+1$, which we then antisymmetrize  
and multiply by $\sqrt{6/N}$.  The factor $1/\sqrt{N}$ guarantees that the typical size of an eigenvalue of $B_{EU}$ will not depend on $N$.  The factor $\sqrt{6}$ has been inserted for later convenience.
Thus $B_{EU} = (\sqrt{6/N})(R - R^T)/2$.  We assume that this matrix is written in a tensor-product basis of the environment 
and the ubit, with the ubit basis being such that the $J_U$ of Eq.~(\ref{wholeS}) has the standard form given in Eq.~(\ref{oneone}).  (A mere rotation of the 
ubit basis does not change $J_U$, but $J_U$ would pick up a minus sign under a reflection.)  For most of our numerical runs the environment dimension $N$ is 200.

To get some insight into how the $EUA$ system will evolve, in the following two subsections we restrict our attention to the simple case in which the system $A$ has no 
dynamics of its own and is not interacting with the ubit.  That is, the local Stueckelbergian $S_{UA}$ is zero,
so that $\hat{S} = S_{EU}\otimes I_A$, where $S_{EU} = -\omega I_E \otimes J_U + s B_{EU}$.  We also assume that 
Alice has prepared the $UA$ system in an initial state that is uncorrelated with the environment, so that
the initial state of the whole system is of the form
\begin{equation}
\hat{\rho}(0) = \rho_E \otimes \rho_{UA}.
\end{equation}
The state $\rho_{UA}$, which could be pure because we are not imposing Stueckelberg's rule, and which may exhibit entanglement between the ubit and the $A$ system, can always be written as
\begin{equation}
\rho_{UA} = I_U \otimes a^{(I)} + J_U \otimes a^{(J)} + X_U \otimes a^{(X)} + Z_U \otimes a^{(Z)},
\end{equation}
where $X_U$ and $Z_U$ are the usual Pauli matrices acting on the ubit's space and the $a$'s are operators on $A$'s space.
With $S_{UA}$ being zero, 
the matrices $\{a^{(I)}, a^{(J)}, a^{(X)}, a^{(Z)}\}$ will not change.  That is, 
at a later time $t$, when the state of the whole system is $\hat{\rho}(t)$, the state of the $UA$ system will be of the form
\begin{equation}
\rho_{UA}(t) = \hbox{Tr}_E\, \hat{\rho}(t) =  I_U \otimes a^{(I)} + u^{(J)}(t) \otimes a^{(J)} + u^{(X)}(t) \otimes a^{(X)} + u^{(Z)}(t) \otimes a^{(Z)}.
\end{equation}
(The ``$I_U$'' part will not change, since $I_U$ commutes with the Stueckelbergian.)   
Our aim here is to follow the evolution of the ubit matrices $u^{(J)}, u^{(X)},$ and $u^{(Z)}$.  

The evolution of these matrices will depend to some extent on the initial state $\rho_E$ of the environment.  Numerically we
have tried both a randomly chosen pure state and the completely mixed state, and we have found that for a sufficiently large 
dimension of the environment's Hilbert space, the results
are almost the same in both cases though they tend to be somewhat smoother in the latter case.  For simplicity, then, in all of our calculations we will choose the initial environment state to be the completely mixed state
$\rho_E = I_E/N$.  

Under these assumptions, our numerical results can be summarized as follows.  The function $u^{(J)}$ is of the form $\gamma(t)J_U$,
where $\gamma(t)$ is an initially oscillating function that finally approaches a constant value between zero and one.  Thus the $J_U$
part of the state diminishes but does not disappear.  On the other hand, the functions $u^{(X)}$ and $u^{(Z)}$ both become 
linear combinations of $X_U$ and $Z_U$ whose coefficient vectors rotate in the $X$-$Z$ plane and finally decay to zero (at least to a very
good approximation, which we expect will be exact in the limit of an infinite-dimensional environment).  
The rotation of the coefficient vectors is simply a manifestation of the rotation of the ubit in its two-dimensional real state space.  That $u^{(X)}$ and $u^{(Z)}$ decay to zero tells us that the state of the $UA$ system eventually obeys
Stueckelberg's rule, since the remaining operators $I_U$ and $J_U$ commute with $J_U$.  These results
can be understood through perturbation theory, as we now show.

\subsection{The function $\gamma(t)$}

The $2 \times 2$ matrix $u^{(J)}$ can be written as
\begin{equation}
u^{(J)}(t) = \frac{1}{N}\hbox{Tr}_E \left[e^{S_{EU}t} (I_E \otimes J_U) e^{-S_{EU}t} \right] .
\end{equation}
As we have said, this matrix remains proportional to $J_U$---this follows
from the fact that $J$ is antisymmetric while $I$, $X$ and $Z$ are symmetric---so that
$u^{(J)}(t) = \gamma(t)J_U$ for some real function $\gamma(t)$.  Here we try to estimate $\gamma(t)$.  We can write it as
\begin{equation}  
\gamma(t) = -\frac{1}{2N}\hbox{Tr} \left[e^{S_{EU}t} (I_E \otimes J_U) e^{-S_{EU}t}(I_E \otimes J_U) \right] .
\end{equation}
To apply perturbation theory, it is helpful to define a Hermitian matrix $G$ as follows:
\begin{equation}
G = \frac{iS_{EU}}{\omega} = -i I_E \otimes J_U + \frac{s}{\omega}(iB_{EU}).
\end{equation}
We can think of $G$ as 
\begin{equation}  \label{one}
G = G_0 + \lambda V, 
\end{equation}
where 
\begin{equation}  \label{two}
G_0 =- i I_E \otimes J_U \hspace{1cm} \hbox{and} \hspace{1cm} V = iB_{EU},
\end{equation}
and $\lambda = s/\omega$ is our perturbation parameter.  
In terms of $G$, we have
\begin{equation} \label{gamma}
\gamma(t) = -\frac{1}{2N}\hbox{Tr} \left[e^{-i\omega Gt} (I_E \otimes J_U) e^{i\omega Gt}(I_E \otimes J_U) \right] .
\end{equation}

Note that $G_0$, which is a $2N \times 2N$ matrix, has only two eigenvalues, $+1$ and $-1$, each associated with 
an $N$-dimensional subspace.  Let $P_+$ and $P_-$ be the projection operators on the subspaces corresponding to
the eigenvalues $+1$ and $-1$ respectively.  We can write these operators explicitly as 
\begin{equation}  \label{twoprojections}
P_+ = \frac{1}{2} I_E \otimes \left(I_U - iJ_U\right) \hspace{1cm} \hbox{and} \hspace{1cm} P_- = \frac{1}{2}I_E \otimes \left(I_U + iJ_U\right).
\end{equation}
To do degenerate perturbation theory, we choose a basis that diagonalizes the matrix $V$ in each
of these two subspaces.  Let $|\Phi_n^+\rangle$ and $|\Phi_n^-\rangle$ be the elements of
this basis.  That is, the vectors $|\Phi_n^+\rangle$ are the eigenvectors of $V^+ = P_+ V P_+$, and the vectors
$|\Phi_n^-\rangle$ are the eigenvectors of $V^- = P_- V P_-$.  Thus $\langle \Phi_n^+| V |\Phi_m^+\rangle$ is zero if $m\ne n$ but $\langle \Phi_n^+| V |\Phi_m^-\rangle$
is typically nonzero.

Before proceeding, it is worth noting certain symmetries that follow from the fact that any Stueckelbergian  
is a real, antisymmetric matrix and that $P_-$ is the complex conjugate of $P_+$.  First, the Hermitian matrix $V^-$ is the negative complex conjugate of $V^+$.  From this it 
follows that we can take $|\Phi_n^-\rangle$ to be the complex conjugate of $|\Phi_n^+\rangle$, and if  $v_n$ is the eigenvalue
of $V^+$
associated with  $|\Phi_n^+\rangle$ then $-v_n$ is the eigenvalue of $V^-$ associated with $|\Phi_n^-\rangle$.
Similarly, the eigenvectors of the Hermitian matrix $G$ can be written as 
$|\Psi_n^+\rangle$ and $|\Psi_n^-\rangle$, corresponding to eigenvalues $g_n$ and $-g_n$ respectively, where $|\Psi_n^-\rangle$ is the complex
conjugate of $|\Psi_n^+\rangle$.    

In the expression (\ref{gamma}) for $\gamma(t)$, we make the substitution
\begin{equation} \label{exponentials}
e^{i\omega G t} = \sum_{n=1}^N \left(e^{i\omega g_n t}|\Psi_n^+\rangle \langle \Psi_n^+| + 
e^{-i\omega g_n t}|\Psi_n^-\rangle \langle \Psi_n^-|\right).
\end{equation}
Leaving the eigenvalues unanalyzed for now, we use standard time-independent perturbation theory to write
the {\em eigenvectors} in terms of the unperturbed eigenvectors $|\Phi_n^+\rangle$ and $|\Phi_n^-\rangle$.  Specifically,
we expand $|\Psi_n^+\rangle$ and $|\Psi_n^-\rangle$ to second order in $\lambda$ (because there is no first-order contribution 
to $\gamma(t)$) and insert this expansion
into Eq.~(\ref{exponentials}), which in turn
is inserted into Eq.~(\ref{gamma}).  The second-order expansion of $|\Psi_n^\pm\rangle$ is given in Appendix A.
The resulting expression for $\gamma(t)$ comes out to be
\begin{equation}  \label{perturbationgamma}
\gamma(t) = 1 - \frac{\lambda^2}{N}\sum_{n,m = 1}^N |\langle \Phi_n^+|V|\Phi_m^-\rangle|^2 \left(1 - \cos\left[(g_n + g_m) \omega t\right]\right).
\end{equation}

With a couple of assumptions about the matrix elements $\langle \Phi_n^+|V|\Phi_m^-\rangle$ and the distribution of values of $g_n$,
we can obtain an explicit functional form for $\gamma(t)$.  First we assume that because many random values are being summed
to get the squared matrix element $|\langle \Phi_n^+|V|\Phi_m^-\rangle|^2$, we can replace this factor, for each value of $m$ and $n$, with its 
ensemble average, that is, the average over all possible matrices $B_{EU}$ generated by the random procedure specified above.  To find
this ensemble average, we
begin with
\begin{equation}  \label{ensave}
\left\langle |\langle \Phi_n^+|V|\Phi_m^-\rangle|^2\right\rangle = \frac{1}{N^2}\sum_{n,m=1}^N \left\langle |\langle \Phi_n^+|V|\Phi_m^-\rangle|^2\right\rangle
= \frac{1}{N^2} \left\langle \hbox{Tr}\left(P_+ V P_- V\right) \right\rangle,
\end{equation}
where the angular bracket indicates the ensemble average.  
Inserting the expressions (\ref{twoprojections}) into Eq.~(\ref{ensave}) and writing $V$ in terms of the random matrix $R$, we find that
each term involving $J_U$ is zero because it is proportional to the ensemble average of the product of two distinct elements of 
the random matrix $R$.  The
remaining terms give us
\begin{equation} \label{earlyapprox}
\left\langle |\langle \Phi_n^+|V|\Phi_m^-\rangle|^2\right\rangle = \frac{1}{4N^2}\left\langle\hbox{Tr}(V^2)\right\rangle= -\frac{6}{16N^3}\left\langle\hbox{Tr}\left[(R^T-R)^2\right]\right\rangle \approx \frac{1}{N},
\end{equation}
where we have used the fact that the average square of a component of $R$ is 1/3, 
and we have neglected $\langle\hbox{Tr}(R^2)\rangle$ and $\langle\hbox{Tr}[(R^T)^2]\rangle$ because they are of order $N$ whereas $\langle\hbox{Tr}(R^TR)\rangle$ is of order $N^2$.  We will use the value $1/N$ in place of $ |\langle \Phi_n^+|V|\Phi_m^-\rangle|^2$ in Eq.~(\ref{perturbationgamma}).
 
 We now turn to the eigenvalues $\pm g_n$ of $G$.  Recall that the unperturbed eigenvalues, that is, the eigenvalues
 of $G_0$, are simply $+1$ and $-1$.  To lowest nontrivial order in $\lambda$, we can write
 $g_n = 1 +\lambda v_n$.  (Again, $v_n = \langle \Phi_n^+|V|\Phi_n^+\rangle$ is an eigenvalue of $V^+$.)  Because $V^+$ comes from linearly transforming a random matrix, for large $N$ we expect its 
 eigenvalues $\{v_n\}$ to approximately exhibit a semicircular distribution.
 To write this distribution explicitly, we find the ensemble average of $(1/N)\sum v_n^2$, reasoning as above but now with a smaller
 matrix:
 \begin{equation}
 \left\langle\frac{1}{N} \sum_n v_n^2 \right\rangle = \frac{1}{N}\left\langle \hbox{Tr}\left[\left( V^+\right)^2 \right] \right\rangle = \frac{1}{4N}\left\langle \hbox{Tr}\left(V^2 \right) \right\rangle \approx 1.
\end{equation}
Let $\eta(v) dv$ be the expected {\em number of eigenvalues} of $V^+$ lying between $v$ and $v+dv$, so that the normalization of $\eta$
is fixed by the condition $\int_{-v_{max}}^{v_{max}} \eta(v) dv = N$, where $v_{max}$ is the maximum value of $v$.  Then the unique semicircular
distribution satisfying $\langle v^2 \rangle = 1$ is given by
\begin{equation}
\eta(v) = (N/\pi)\sqrt{1-(v/2)^2},
\end{equation}
so that $v_{max} = 2$.
To get our analytic expression for $\gamma(t)$, we replace $g_n$
in Eq.~(\ref{perturbationgamma}) with $1+\lambda v$ and $g_m$ with $1 + \lambda v'$, and we assume both $v$ and $v'$
are distributed according to $\eta$.  
 
\begin{figure}[h]
\begin{center}
\includegraphics[scale = 1]{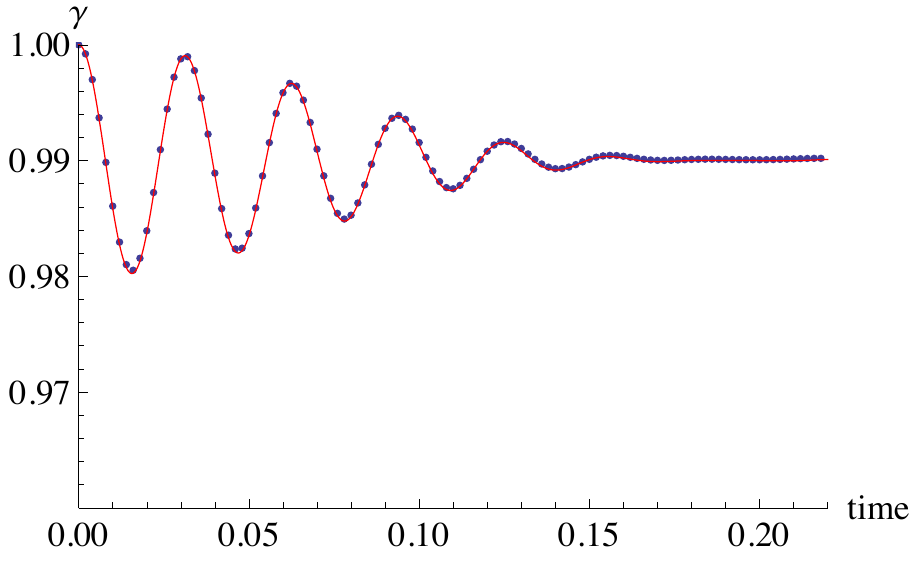}
\end{center}
\vspace{-3mm}
\caption{(Color online) The evolution of the coefficient $\gamma(t)$ of $J_U$ under the influence of the environment.  The initial value of $\gamma$
is set to unity, and the system is initially in a product state, with no correlation between the ubit and the environment.
The dots (blue) show the numerical results, and the curve (red) shows our result from perturbation theory 
(Eq.~(\ref{gammafinal})).  Here $N=200$, $s = 10$ and $\omega = 100$.  (Time is measured in arbitrary units.  If 
$t$ is interpreted as being in zs, for example, 
then $s = 10$ zs$^{-1}$ and $\omega = 100$ zs$^{-1}$.)  In this figure, as in every figure in this paper,
the numerical results are for a single run of the simulation.  The results from one run to the next are extremely consistent.}
\label{gammafigure}
\end{figure}

 With these approximations, we have
 \begin{equation}
 \gamma(t) = 1 - \lambda^2 + \frac{\lambda^2}{N^2} \int_{-2}^{2} \int_{-2}^{2} 
 \eta(v)\eta(v') \cos\left[ \left( 2 + \lambda v + \lambda v' \right) \omega t \right] dv\, dv'.
 \end{equation}
 The integral can be done exactly, and we obtain
 \begin{equation}  \label{gammafinal}
 \gamma(t) = 1 - \lambda^2 + \left[\frac{J_1(2{s} t)}{\omega t}\right]^2
 \cos(2\omega t),
 \end{equation}
 where $J_1$ is the Bessel function of order 1; that is,
 \begin{equation}
 J_1(x) = \sum_{n=0}^\infty \frac{(-1)^n}{n! (n+1)!} \left( \frac{x}{2} \right)^{2n+1}.
 \end{equation}
 When $t$ is large, $\gamma(t)$ approaches the constant value $1 - \lambda^2$.  That is, the $J$ part of 
 the matrix $\rho_{UA}$ is reduced but does not vanish.  This analytic expression is compared with our numerical results
 in Fig.~\ref{gammafigure}.  
 
 \subsection{The matrix $u^{(X)}$}

We would now like to get an analytic expression for $u^{(X)}$, which starts out as $X$ at $t=0$.  The matrix $u^{(X)}$ can be written as 
\begin{equation}  \label{uX}
u^{(X)}(t) = \frac{1}{N}\hbox{Tr}_E \left[e^{S_{EU}t} (I_E \otimes X_U) e^{-S_{EU}t} \right].
\end{equation}
We can always write a $2 \times 2$ matrix as a linear combination of basis elements, so we can write $u^{(X)}$ as
\begin{equation}
u^{(X)}(t) = \beta^{(I)}(t) I+ \beta^{(J)}(t) J+ \beta^{(X)}(t) X + \beta^{(Z)}(t) Z ,
\end{equation}
where the $\beta$'s are real-valued functions.
As $X$ is symmetric, $\beta^{(J)}$ must equal zero since $J$ is anti-symmetric.  We also know that $\beta^{(I)}$ equals zero since the identity commutes with $e^{S_{EU}t}$:
\begin{equation}
\beta^{(I)} = \frac{1}{2 N}\hbox{Tr}\left[ (I_E \otimes I_U) e^{S_{EU}t} (I_E \otimes X_U) e^{-S_{EU}t}  \right] = \frac{1}{2 N}\hbox{Tr}\left[ I_E \otimes X_U \right] = 0.
\end{equation}
Thus $u^{(X)}(t) = \beta^{(X)}(t) X + \beta^{(Z)}(t) Z$.

To calculate $\beta^{(X)}$ and $\beta^{(Z)}$, we again apply perturbation theory.  We begin with the following expression for $\beta^{(X)}$:
\begin{equation} \label{betaX}
\beta^{(X)}(t) = \frac{1}{2 N}\hbox{Tr}\left[ (I_E \otimes X_U) e^{S_{EU}t} (I_E \otimes X_U) e^{-S_{EU}t}  \right].
\end{equation}
When writing $e^{S_{EU}t}$ in terms of the unperturbed eigenvectors, it is convenient to factor each unperturbed eigenvector into a tensor product of the environment and ubit parts.  The eigenvectors of $-iJ$ (the ubit part of $G_0$) are $|+\rangle = \frac{1}{\sqrt{2}} \bigl( \begin{smallmatrix} 1\\ -i \end{smallmatrix} \bigr)$ and $|-\rangle = \frac{1}{\sqrt{2}} \bigl( \begin{smallmatrix} 1\\ i \end{smallmatrix} \bigr)$, corresponding to
the eigenvalues $+1$ and $-1$.  Thus, 
\begin{equation}  \label{firstphi}
 | \Phi_n^+ \rangle = | \phi_n^+ \rangle \otimes |+\rangle
\end{equation}
\begin{equation}  \label{secondphi}
 | \Phi_n^- \rangle = | \phi_n^- \rangle \otimes |-\rangle
\end{equation}
for some environment states $|\phi_n^\pm\rangle$ (where $|\phi_n^-\rangle$ is the complex conjugate of $|\phi_n^+\rangle$).
We now write $X$ in terms of $|+\rangle$ and $|-\rangle$:
\begin{equation} \label{X}
X = \mtx{ccc}{0& &1\\1& &0} = \hbox{i}\left( |+\rangle \langle - | - |-\rangle \langle + | \right).
\end{equation}
Inserting Eqs.~(\ref{exponentials}) and (\ref{X}) into Eq.~(\ref{betaX}), we find an expression for $\beta^{(X)}$ to zeroth order in $\lambda$ for the eigenstates.  There are no first order terms, and the zeroth-order terms turn out to be sufficient to give us good agreement with
the numerical calculations.  The resulting expression comes out to be
\begin{equation} \label{bxt}
\beta^{(X)}(t)=  2 \sum_{n,m = 1}^N |\langle \phi_n^+|\phi_m^-\rangle|^2 \cos\left[(g_n + g_m) \omega t\right].
\end{equation}

\begin{figure}[h]
\begin{center}
\includegraphics[scale = 0.85]{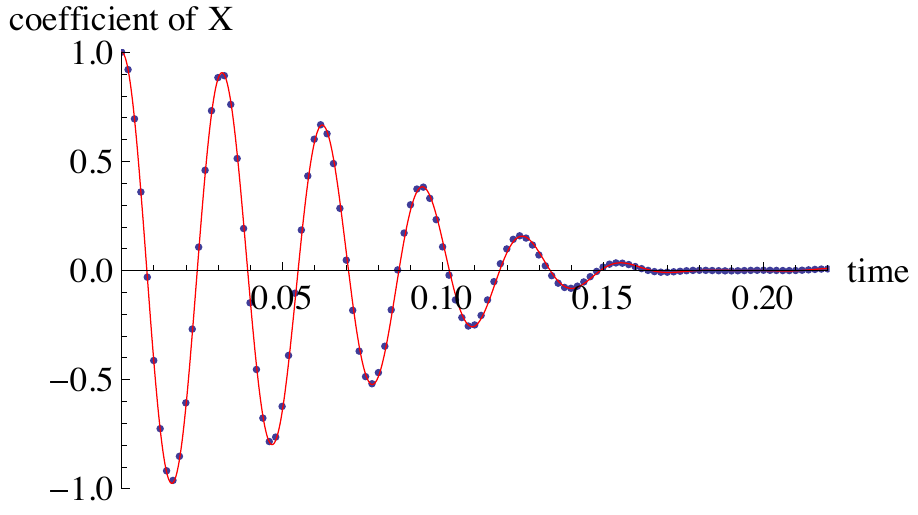}
\includegraphics[scale = 0.85]{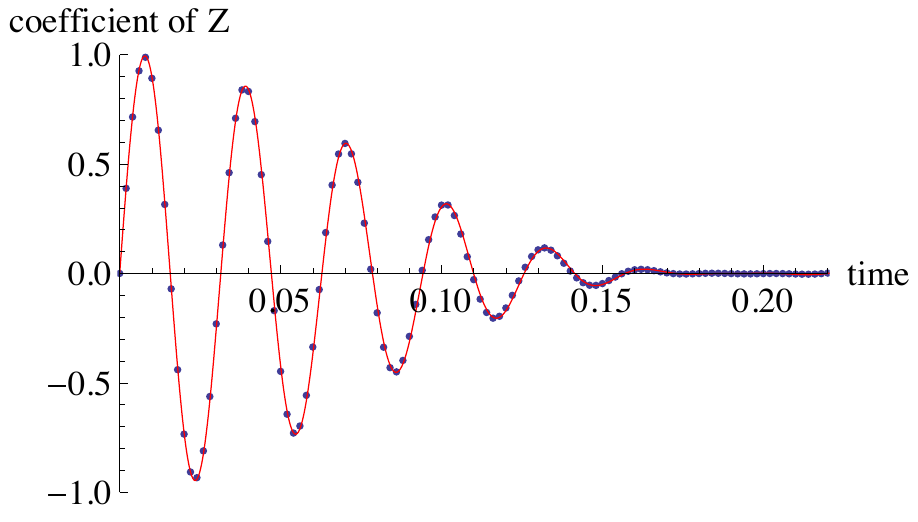}
\end{center}
\vspace{-3mm}
\caption{(Color online) Evolution of the matrix $u^{(X)}(t) = \beta^{(X)}X + \beta^{(Z)}Z$, which begins as $X$.  Here we plot $\beta^{(X)}$ and $\beta^{(Z)}$ as functions
of time.  One can see
that the vector $(\beta^{(X)}, \beta^{(Z)})$ rotates in the plane and finally decays to zero. The dots (blue) show the numerical
results, and the curves (red) show
the result of our perturbation theory calculation (Eqs.~(\ref{betaXfinal}) and (\ref{betaZ})).  Here $N=200$, $s = 10$ and $\omega = 100$.
As in Fig.~1, the time is in arbitrary units, with corresponding units for $s$ and $\omega$.}
\label{betafigure}
\end{figure}

Using the same assumption as before about the distribution of the values of $g_n$, we can 
find an expression for $\beta^{(X)}$ independent of the details of the eigenstates and eigenvalues of $G$.  We also assume that for a sufficiently large environment, we can approximate each $|\langle \phi_n^+ | \phi_m^-\rangle|^2$ by its ensemble average $1/N$.
Plugging this value back into Eq.~(\ref{bxt}) and assuming the same semicircle distribution as before, we arrive at the final expression:
\begin{equation}  \label{betaXfinal}
	\beta^{(X)}(t) = \left[\frac{J_1(2{s} t)}{{s}t}\right]^2   \cos(2 \omega t).
\end{equation} 
In a similar way, we find that 
\begin{equation}  \label{betaZ}
\beta^{(Z)}(t) = \frac{1}{N} \sum_{n,m = 1}^N |\langle \phi_n^+|\phi_m^-\rangle|^2 \sin\left[(g_n + g_m) \omega t\right] = \left[\frac{J_1(2{s} t)}{{s}t}\right]^2  \sin(2 \omega t).
\end{equation}
The functions $\beta^{(X)}$ and $\beta^{(Z)}$ together show what happens as $u^{(X)}$ evolves over time.  The $X$ and $Z$ parts of the ubit's matrix $u^{(X)}$ rotate into each other and eventually decay to zero.  Fig.~\ref{betafigure} compares our analytic expressions for 
both $\beta^{(X)}$ and $\beta^{(Z)}$ to our numerical results.

Solving for $u^{(Z)}$ yields a similar result.  The ubit matrix $u^{(Z)}$ begins as $Z$, and the $Z$ and $X$ parts again rotate into each other as they decay down to zero.  

\subsection{Projecting onto the space of matrices that commute with $S_{EU}$}

The above calculations show that $\gamma(t)$, $\beta^{(X)}(t)$, and $\beta^{(Z)}(t)$ all approach asymptotic values.
We could have obtained these asymptotic values more directly in the following way.
First, we can break $\hat{\rho}(0)$ into two parts: the part $\hat{\rho}_\parallel$ lying in the space of matrices 
that commute with $S_{EU} \otimes I_A$, and the part $\hat{\rho}_\perp$ perpendicular to this space \footnote{We say two 
matrices $M_1$ and $M_2$ are orthogonal when $\hbox{Tr}\,M_1^\dag M_2 = 0$.}.  Then the evolution (again in the 
absence of any local Stueckelbergian $S_{UA}$) becomes
\begin{equation}
\hat{\rho}(t) = e^{(S_{EU} \otimes I_A) t} (\hat{\rho}_\parallel + \hat{\rho}_\perp) e^{-(S_{EU} \otimes I_A) t} = \hat{\rho}_\parallel 
+ e^{(S_{EU} \otimes I_A) t} \hat{\rho}_\perp e^{-(S_{EU} \otimes I_A) t}.
\end{equation}
The second term is what provides the oscillations we saw in the graphs in the preceding subsections.  Asymptotically the effect of 
this second term on the $UA$ system disappears as the oscillating components tend to cancel each other out when we 
trace over the environment.  It is only
$\hat{\rho}_\parallel$ that contributes to the final state of $UA$; so in order to find that state we could have focused only on
$\hat{\rho}_\parallel$ (to which we could have applied perturbation theory as in the above calculations).  

For the remainder of this paper, in our analytical work we will adopt the following ansatz.  First, we assume that both
$s$ and $\omega$, that is, the scales of the two terms in the $EU$ part of the Stueckelbergian, are much larger in magnitude
than the spread in eigenvalues of the local Stueckelbergian $S_{UA}$.  We imagine taking the limit as $s$ and $\omega$
both go to infinity while their ratio remains fixed.  As $s$ and $\omega$ get larger, the oscillations and decay we observed
in the preceding subsections simply proceed at a faster rate without changing their form in any other way.  
Consider, then, the evolution of $\hat{\rho}$ over any short interval of time.  By the time $S_{UA}$ has had any appreciable effect, the asymptotic value of $\rho_{UA}$ due to the action of $S_{EU}$ will have already been reached.  Therefore, for the purpose of computing
$\rho_{UA}$ we will (i) ignore any initial $\hat{\rho}_\perp$ and (ii) assume that $\hat{\rho}_\parallel$, as it evolves, is continually projected
into the space of matrices that commute with $S_{EU}\otimes I_A$.  That is, we will use this continual projection in place
of the exact evolution due to $S_{EU}$ in all our later analytical calculations.  (But in our numerical work we will follow the 
exact dynamics.)  
Note that this method of continual projection does not necessarily provide a good approximation to $\hat{\rho}$ itself.  The rapidly oscillating part 
$\hat{\rho}_\perp$ does not go away, but it is irrelevant for computing $\rho_{UA}$.  To remind ourselves that we are dealing only with
$\hat{\rho}_\parallel$ and not with the full density matrix $\hat{\rho}$, we will keep the subscript ``$\parallel$'' when referring to the state of 
the whole system.  
In the following paragraphs we 
write down some general consequences of the assumption of continual projection.  

First we modify the equation of evolution, Eq.~(\ref{diffeq}), so that it does not allow
$\hat{\rho}_\parallel$ to evolve away from the space of matrices that commute with $S_{EU}\otimes I_A$.  The modified equation is
\begin{equation}  \label{projdiffeq}
\frac{d\hat{\rho}_\parallel}{dt} = {\mathcal P}\left( \left[ I_E \otimes S_{UA}, {\mathcal P}(\hat{\rho}_\parallel)\right] \right),
\end{equation}
where ${\mathcal P}$ projects onto this space.  (The second ${\mathcal P}$ is unnecessary, since $\hat{\rho}_\parallel$ is already 
in the space into which ${\mathcal P}$ projects.  We include it only so that the equation would preserve the trace of {\em any} 
density matrix.)  Assuming that $S_{EU}$ is non-degenerate, the only matrices that commute with it
are linear combinations of projections onto its eigenstates.   Thus we can express the action of ${\mathcal P}$ on a generic
matrix $M$ as follows:
\begin{equation} \label{proj}
{\mathcal P}(M) = \sum_j |\Psi_j\rangle\langle\Psi_j| \otimes \hbox{Tr}_{EU}\left[\left(|\Psi_j\rangle\langle\Psi_j|\otimes I_A\right)M\right].
\end{equation}
Here we use the index $j$ to stand for the combination of $n$ and $\pm$ in $|\Psi_n^\pm\rangle$.  (Again, the vectors
$|\Psi_n^\pm\rangle$ are the eigenvectors of $S_{EU}$.)  
The most general form of $\hat{\rho}_\parallel$ as a function of time (still assuming that $S_{EU}$ is nondegenerate) is 
\begin{equation}  \label{projrho}
\hat{\rho}_\parallel(t) = \frac{1}{2N}\sum_j |\Psi_j\rangle\langle\Psi_j| \otimes \sigma_j(t),
\end{equation}
where $\sigma_j(t)$ is a matrix acting on the space of the $A$ system.  Inserting Eqs.~(\ref{proj}) and (\ref{projrho}) into Eq.~(\ref{projdiffeq}),
we get the following equation for the evolution of the $\sigma_j$'s.
\begin{equation}  \label{basicAequation}
\frac{d\sigma_j}{dt} = \left[ \langle\Psi_j|I_E \otimes S_{UA} |\Psi_j\rangle, \sigma_j\right],
\end{equation}
where the quantity $\langle\Psi_j|I_E \otimes S_{UA} |\Psi_j\rangle$ is to be interpreted as a ``partial expectation value,'' in which
the $EU$ vector $|\Psi_j\rangle$ combines with the $EU$ part of $I_E \otimes S_{UA}$ to leave a matrix that acts on the space
of the $A$ system.  

Now, the local Stueckelbergian $S_{UA}$ must be a linear combination of the four ubit matrices $I_U$, $J_U$, $X_U$, and $Z_U$, each in a tensor
product with some matrix of the $A$ system.  Eq.~(\ref{basicAequation}) thus calls on us to evaluate the quantities
$\langle\Psi_j|I_E \otimes I_U |\Psi_j\rangle$, $\langle\Psi_j|I_E \otimes J_U |\Psi_j\rangle$, $\langle\Psi_j|I_E \otimes X_U |\Psi_j\rangle$, 
and $\langle\Psi_j|I_E \otimes Z_U |\Psi_j\rangle$. The first of these is clearly equal to unity.  
If $|\Psi_j\rangle$ were simply a random state, then the other three would have typical values that diminish in magnitude proportional to $1/\sqrt{N}$ for large $N$ \cite{Lubkin}.  
But $|\Psi_j\rangle$ is not a random state.  Rather, it is an eigenstate of $S_{EU} = -\omega I_E \otimes J_U + s B_{EU}$ where 
$B_{EU}$ is random.  The presence of $\omega I_E \otimes J_U$ in this matrix prevents $\langle\Psi_j|I_E \otimes J_U |\Psi_j\rangle$ from
going to zero as $N$ approaches infinity, but it does not similarly protect $\langle\Psi_j|I_E \otimes X_U |\Psi_j\rangle$ or
$\langle\Psi_j|I_E \otimes Z_U |\Psi_j\rangle$.  We find numerically that even with a nonzero value of $\omega$ these last two quantities have typical values that approach
zero as $1/\sqrt{N}$, while the typical size of $\langle\Psi_j|I_E \otimes J_U |\Psi_j\rangle$ approaches a nonzero constant.   
Let us define 
\begin{equation}  \label{nudef}
\nu_j = -i\langle\Psi_j|I_E \otimes J_U |\Psi_j\rangle,
\end{equation}
in which the $i$ has been inserted to make $\nu_j$ real.  In Section V we will estimate $\nu_j$, but for now we simply use it to 
rewrite our basic equation (\ref{basicAequation}).  According to what we have just said, in the large $N$ limit we can ignore any part of $S_{UA}$ that is
proportional to $X_U$ or $Z_U$, so that in effect the most general $S_{UA}$ has the form
\begin{equation}  \label{effectiveS}
S_{UA} = I_U \otimes L_A - J_U \otimes K_A,
\end{equation}
where $L_A$ is an antisymmetric real matrix and $K_A$ is a symmetric real matrix.
Inserting this form into Eq.~(\ref{basicAequation}), we get
\begin{equation}  \label{pseudoH}
\frac{d\sigma_j}{dt} = -i[\nu_jK_A + iL_A, \sigma_j].
\end{equation}
(The matrix $\sigma_j$ can be complex, as long as the imaginary parts cancel out when we 
do the sum in Eq.~(\ref{projrho}).)  
Evidently the Hermitian matrix $\nu_jK_A + iL_A$ is playing a role like that of $H/\hbar$, except that because of the 
$j$ dependence in $\nu_j$, different 
components $\sigma_j$ can have different effective Hamiltonians.  We will get a sense of what consequences this fact has
as we consider in Sections IV and V the special case of a precessing spin.  

The fact that $\langle\Psi_j|I_E \otimes X_U |\Psi_j\rangle$ and $\langle\Psi_j|I_E \otimes Z_U |\Psi_j\rangle$ become zero in the 
large $N$ limit has another important consequence.  First, it means that when we expand $|\Psi_j\rangle\langle\Psi_j|$ as a linear
combination of the ubit matrices $I_U$, $J_U$, $X_U$, and $Z_U$, the environment matrices multiplying $X_U$ and $Z_U$
must have zero trace.  Therefore, for any $\hat{\rho}_\parallel$ of the form given in Eq.~(\ref{projrho}), the density matrix $\rho_{UA}$
resulting from tracing over the environment cannot include any term proportional to $X_U$ or $Z_U$ (in the limit as
$N$ approaches infinity), so that $\rho_{UA}$ commutes with $J_U$. Thus both our local Stueckelbergian $S_{UA}$ and our
local density matrix $\rho_{UA}$ commute with $J_U$, and in this sense we have recovered Stueckelberg's rule 
through the interaction of the ubit with the large environment.  We have not explicitly ruled out the possibility of a {\em measurement}
operator
that does not commute with $J_U$, but if Alice were to manage
to perform such a measurement, represented by a projection operator $P_{UA}$, the anticommuting part 
$(P_{UA})_{a}=(1/2)[P_{UA} + (J_U\otimes I_A)P_{UA} (J_U\otimes I_A)]$ 
would make no observable difference because
$\hbox{Tr}\big[\rho_{UA}(P_{UA})_{a}\big]$ is equal to zero for any $\rho_{UA}$ that commutes with $J_U \otimes I_A$.
(In principle the measurement {\em could}, as a result of the projection $P_{UA}$, create a state $\rho_{UA}$ that does not commute with $J_U \otimes I_A$, but
in the limit we are considering the noncommuting part would immediately decay to zero.)  
We hasten to add, though, that this effective enforcement of Stueckelberg's rule by our projection hypothesis 
does not make our theory equivalent to standard quantum 
mechanics.  It eliminates unwanted states and unwanted terms in the Stueckelbergian, but as we will see, it is
not equivalent to simply imposing Stueckelberg's rule as in Section II.  The interaction with the environment yields a different 
effective dynamics.
Our main task in the rest of this paper is to characterize the differences.

Finally, one might wonder whether it really makes sense to assume, as we have done, that the parameter $s$, which is the size
of a typical eigenvalue of $s B_{EU}$, is much larger 
in magnitude than the spread in eigenvalues
of the local Stueckelbergian.  After all, the idea underlying our model is that the ubit's interaction with each component of the rest of the world  
should be similar to its interaction with the local system.  

In fact there is no contradiction here.  In a more realistic model of 
the environment, the size of a typical eigenvalue of the ubit-environment Stueckelbergian would reflect not just the strength of interaction
between the ubit and a single component of the environment.  It would also reflect the size of the environment.  Suppose, for example,
that the environment consists of $n$ rebits and that the ubit interacts with each one via a simple $4\times 4$ Stueckelbergian matrix
with eigenvalues $\{i\xi, i\xi, -i\xi, -i\xi\}$.  Then even if those individual rebits do not interact with each other, the square root of the average
squared magnitude of an eigenvalue of the whole interaction Stueckelbergian is equal to $\xi\sqrt{n}$.  Thus the typical size
of an eigenvalue grows with the size of the environment.  That is, if we were to write the ubit-environment Stueckelbergian as
$sB_{EU}$, with $B_{EU}$ scaled so that the typical size of its eigenvalues is independent of the size of the environment (as
in Subsection III.A), then the value of 
$s$ would have to grow with the environment.  To be sure, this model of the environment as composed of independent systems
is not the one we have chosen for our numerical simulations, but this argument shows that it is reasonable to assume that $s$ is large: it is large by virtue of the large size of the environment, even if the strength
of interaction between the ubit and any small component of the environment is of limited magnitude.  

\subsection{No signaling}

To summarize the last subsection: by imagining both $s$ and $\omega$ going to infinity with a fixed ratio, we were led to 
consider only the part of the global density matrix that commutes with $S_{EU} \otimes I_A$, and we assumed that during the evolution, this 
part is continually projected onto the space of matrices that commute with $S_{EU} \otimes I_A$.  This assumption led us
to the form (\ref{projrho}) of the density matrix, which evolves according to Eq.~(\ref{basicAequation}).  Next, we considered
the implications of the environment's dimension becoming infinitely large.  (In our model, this means that the number of
independently chosen random parameters becomes infinitely large.)  In this limit, we concluded---admittedly on the basis
of numerical evidence---that 
certain terms of the local Stueckelbergian will become inconsequential, because the random nature of the matrix $B_{EU}$
causes the contributions of these terms to Eq.~(\ref{basicAequation}) to vanish.  The only terms in $S_{EU}$ that can have any effects, then,
are those
that commute with $J_U$.  In that case Eq.~(\ref{basicAequation}) can be written in the specific form (\ref{pseudoH}).  
We now show that this form of the equation does not allow signaling between two observers Alice and Bob if the systems they hold,
$A$ and $B$,
have no direct or indirect interaction between them except through the ubit.  

We first have to write down what it means
that $A$ and $B$ are not interacting except through the ubit.  In standard quantum theory, two isolated and therefore non-interacting 
systems ${\mathcal A}$
and ${\mathcal B}$ have a Hamiltonian of the form
\begin{equation}
H_{{\mathcal A}{\mathcal B}} = H_{\mathcal A}\otimes I_{\mathcal B} + I_{\mathcal A} \otimes H_{\mathcal B}.
\end{equation}
(We use script letters to refer to complex-vector-space systems.)  
Converting this Hamiltonian to real-vector-space language as in Section II, we have that the Stueckelbergian is
\begin{equation}
\hbar S_{UAB} = I_U \otimes \hbox{Re}\left(-iH_{\mathcal A}\otimes I_{\mathcal B} -i I_{\mathcal A} \otimes H_{\mathcal B}\right)
+ J_U \otimes \hbox{Im}\left(-iH_{\mathcal A}\otimes I_{\mathcal B} -i I_{\mathcal A} \otimes H_{\mathcal B}\right).
\end{equation}
We can write this operator in a form like that of Eq.~(\ref{effectiveS}):
\begin{equation}  \label{noninteracting}
S_{UAB} = (I_U \otimes L_A - J_U \otimes K_A) \otimes I_B + I_A \otimes (I_U \otimes L_B - J_U \otimes K_B).
\end{equation}
Given that we are ruling out any terms proportional to $X_U$ or $Z_U$, the form given in Eq.~(\ref{noninteracting})
is the most general form possible for a pair of isolated systems (that is, isolated except for their interaction with the ubit).  

For the pair $AB$, we can rewrite Eq.~(\ref{projrho}) as
\begin{equation}
\hat{\rho}_\parallel(t) = \frac{1}{2N}\sum_j |\Psi_j\rangle\langle\Psi_j| \otimes \tau_j(t),
\end{equation}
where $\tau_j$ is an operator on the space of the $AB$ system, and the hat here labels
an operator on the whole $EUAB$ system.  With a Stueckelbergian of the form (\ref{noninteracting}), the equation of evolution for 
$\tau_j$ is a modified form of Eq.~(\ref{pseudoH}):
\begin{equation}
\frac{d\tau_j}{dt} = -i[(\nu_jK_A + iL_A)\otimes I_B + I_A\otimes(\nu_jK_B + iL_B), \tau_j].
\end{equation}
We are now in familiar territory.  The matrix $\tau_j$ acts on the space of $A$ and $B$, but it evolves according to an effective
Hamiltonian that includes no interaction between the two systems.  Therefore the partial trace of $\tau_j$ over either
of two systems evolves under its own effective Hamiltonian, with no influence from the other system.  E.g., 
\begin{equation}
\frac{d}{dt} \hbox{Tr}_B\, \tau_j = -i\left[ \nu_j K_A + iL_A,  \hbox{Tr}_B\, \tau_j \right].
\end{equation}
It follows that $\rho_{UA}$, which is 
\begin{equation}
\rho_{UA} = \hbox{Tr}_{EB}\, \hat{\rho}_\parallel =\frac{1}{2N} \sum_j \hbox{Tr}_E \left( |\Psi_j\rangle\langle \Psi_j | \right) \hbox{Tr}_B \left( \tau_j(t) \right),
\end{equation}
evolves independently of what happens to system $B$.  
That is, Bob's choice of local Stueckelbergian $S_{UB} = I_U \otimes L_B - J_U \otimes K_B$ cannot affect what Alice sees.  
Now, Bob can perform operations other than simply applying a Stueckelbergian to the $UB$ system for a period of time.  
He can allow system $B$ to interact with other systems (which by assumption are not 
interacting with $A$) and in particular he can make measurements.  But any such operation can be accounted for simply by expanding the definition of system $B$.  (Bob may observe a definite outcome of a measurement, but he is not allowed to convey to Alice any 
information about this outcome.  So
to describe the state Alice observes we need to keep all the outcomes, and we can do this
by letting $B$ become entangled with the measuring device with no collapse.)  We conclude that
in this model, in the limiting case we are considering, there can be no signaling through the ubit.  

To be sure, if $s$, $\omega$, and $N$ remain finite, then there {\em can} be signaling through the ubit.  It would be interesting
to determine quantitatively how the degree of signaling (suitably defined) depends on 
$s$, $\omega$, and $N$, but we leave this question for future work.  Here we focus on the limiting case.  

\begin{figure}[h]
\begin{center}
\includegraphics[scale = 0.8]{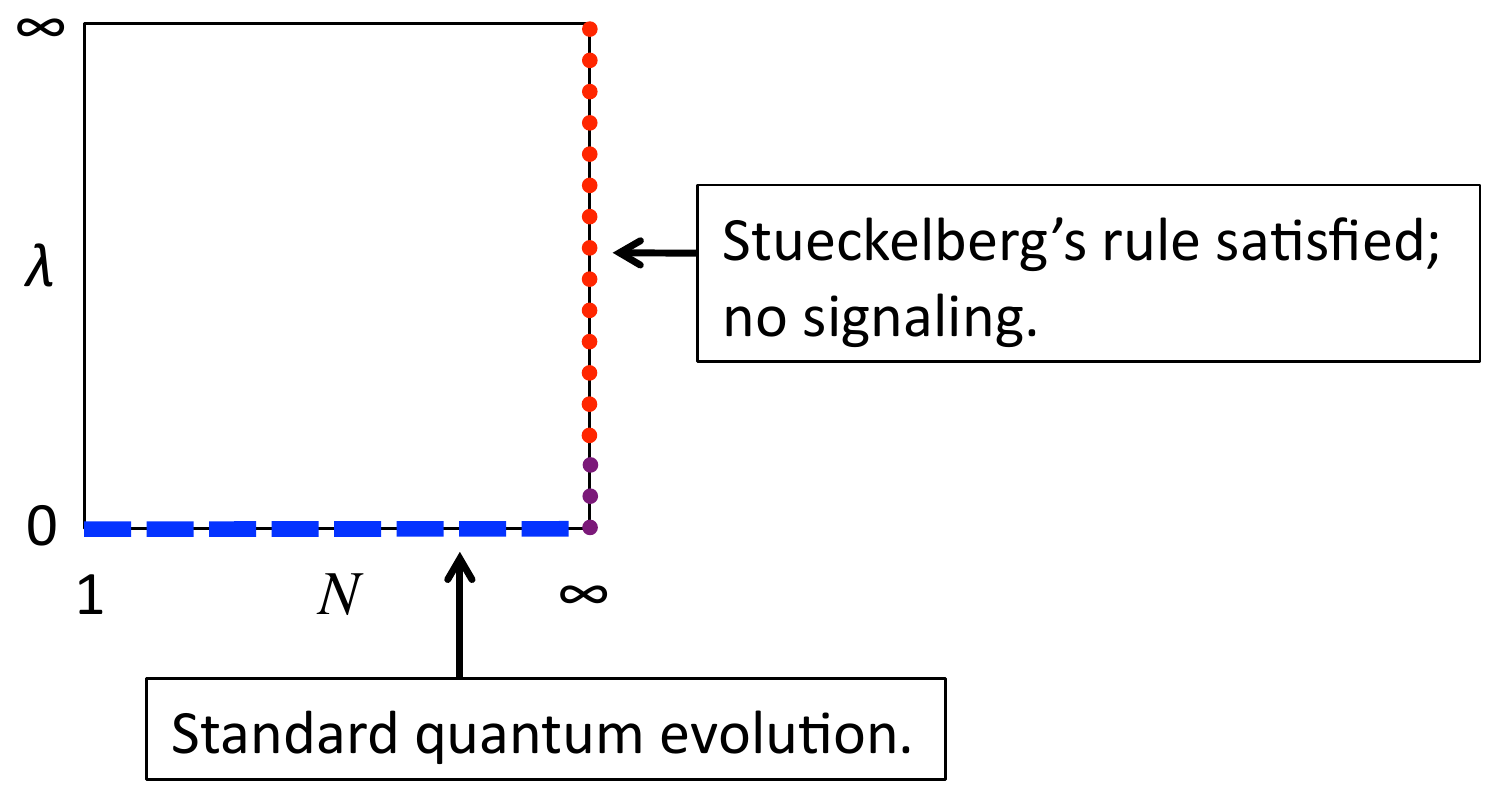}
\end{center}
\vspace{-3mm}
\caption{(Color online) A schematic diagram showing the limits we consider in this paper.  The axes are $N$ (the dimension of the
environment) and $\lambda = s/\omega$.  ($N=1$ corresponds to the case of no environment at all.)  Each point in the diagram
corresponds to a limiting case of infinite $\omega$, but the limits are taken in different ways as we now explain. 
Most of our work in this paper addresses the range 
corresponding to the
vertical dotted line (red and purple).  For each of these points, $s$ and $\omega$ have been taken to infinity with a fixed ratio $\lambda$.  In this regime Stueckelberg's rule is satisfied and there is no signaling, but the theory differs 
from standard quantum theory except as one reaches the bottommost point.  The few lowest dots of that line (purple) are intended to indicate
a small region in which the theory---to the extent we have developed it---is consistent with experiment (see Section VIII for a discussion).  The 
horizontal dashed line (blue) represents the range of cases considered in Appendix B, in which $\omega$ approaches infinity 
with $s$ and $N$ held fixed.  There the evolution operator
is equivalent to the standard quantum mechanical evolution operator, but for those cases we have not proved that 
the states and measurement operators must be equivalent to those of standard quantum theory.  Thus the only route to standard quantum
theory that we claim on this diagram is to follow the vertical dotted line to its lowest point.}
\label{schematicfigure}
\end{figure}

Thus for any fixed value of the ratio $s/\omega = \lambda$, we should get an effective theory that is a no-signaling theory.  
The effective theory will typically not be the same as quantum mechanics---it will be a modification of quantum mechanics.  
The results of the next two sections indicate that for the special example we consider there---a precessing spin---we
recover standard quantum theory when $\lambda$ goes to zero but encounter deviations from quantum theory whenever $\lambda$ is non-zero.  In Appendix B we present
a related argument that does not depend on taking either $s$ or $N$ to be infinite and that applies 
to a general system.  There we show that with fixed $s$ and $N$, 
the evolution operator $e^{\hat{S}t}$ becomes equivalent to the standard quantum mechanical operator when $\omega$ goes to
infinity.  However, the same argument strongly suggests that there will be no such equivalence for any finite value of $\omega$.  
Note also that we consider in Appendix B only the evolution operator; we have
not shown that in the limit $\omega \rightarrow \infty$ (with fixed $s$ and $N$), the states and measurement operators of the ubit model become equivalent to those of standard quantum theory.  
Fig.~\ref{schematicfigure} indicates schematically the limits in which we recover various aspects of 
standard quantum theory.

\section{A Precessing Spin---Numerical Simulations}

\subsection{The Stueckelbergian and the initial state}

In the numerical work of the preceding section we considered the interaction of the ubit with the environment, in the absence of any interaction with the local system
$A$.  We now add this interaction and study the simplest possible case, a precessing qubit.  For definiteness we take the qubit to be
the spin of a spin-1/2 particle in the presence of a constant magnetic field $\vec{B}$.  With $\vec{B}$ in the positive $z$ direction and the particle having a negative charge, the usual Hamiltonian can be written as
$H = \hbar (\Omega/2) Z$,
where $\Omega$ is the angular precession frequency, equal to $B$ times the magnitude of the particle's gyromagnetic ratio, and $Z$ is 
again the Pauli matrix for the $z$ axis.  The precession will be in the right-hand sense around the positive $z$ axis.  

We can use the correspondence given in Section II to re-express the same phenomenon in terms of a rebit $A$ and the ubit.  
The Stueckelbergian is obtained from $H$ as in Eq.~(\ref{HtoS}):
\begin{equation}
S_{UA} = J_U \otimes \hbox{Im}(-iH/\hbar) = -\frac{\Omega}{2}J_U \otimes Z_A.
\end{equation}
Let the initial spin state be in the $x$ direction; that is, the initial density matrix of the qubit is
$\rho = (1/2)(I + X)$.  To re-express this state in the ubit model, we use Eq.~(\ref{complextoreal}):
\begin{equation}
\rho_{UA}(0) = \frac{1}{4}I_U \otimes \left(I_A +  X_A\right).
\end{equation}
If we were to let the $UA$ system evolve under the Stueckelbergian $S_{UA}$ from the starting state $\rho_{UA}(0)$, 
it would exhibit the standard precession simply re-expressed in real-vector-space terms.  That is, we would have
\begin{equation}  \label{standardevol}
\rho_{UA}(t) = \frac{1}{4}\left[ I_U \otimes I_A +  \cos(\Omega t) I_U \otimes X_A + \sin(\Omega t) J_U \otimes J_A \right].
\end{equation}
The combination $J_U \otimes J_A$ takes the place of the Pauli matrix $Y$.  

But we are interested in the evolution of the state of the whole system, with the initial state
\begin{equation}  \label{wholeinitialstate}
\hat{\rho}_\parallel(0) = \frac{1}{4N} I_E \otimes I_U \otimes \left(I_A + X_A \right),
\end{equation}
under the full Stueckelbergian 
\begin{equation}  \label{precessionS}
\hat{S} = -\omega I_E \otimes J_U \otimes I_A + s B_{EU} \otimes I_A + I_E \otimes S_{UA}.
\end{equation}
(As in Section III, we are taking the initial state of the environment to be the completely mixed state.  We make this choice as a matter of simplicity.  Note
that Eq.~(\ref{wholeinitialstate}) is a special case of Eq.~(\ref{projrho}).)
With this initial state and Stueckelbergian, we find the 
state of the $UA$ system at time $t$ by computing
\begin{equation}  \label{exactrhoUA}
\rho_{UA}(t) = \hbox{Tr}_E \left( e^{\hat{S}t}\hat{\rho}_\parallel(0)e^{-\hat{S}t}\right).
\end{equation}
The results of Section III lead us to expect that any component of $\rho_{UA}(t)$ involving $X_U$ or $Z_U$ will be extremely small,
and indeed this is what we observe numerically---those components seem to approach zero as $N$ increases.  This leaves
four basic symmetric matrices in which we can expand $\rho_{UA}(t)$.  The expansion can be written as
\begin{equation}
\rho_{UA}(t) = \frac{1}{4}\left[I_U \otimes I_A + b_x I_U \otimes X_A + b_y J_U \otimes J_A + b_z I_U \otimes Z_A\right].
\end{equation}
Thus in our model we can still imagine the evolution of the spin as a path through the Bloch sphere, with the Bloch vector defined 
as $\vec{b} = (b_x, b_y, b_z)$.  This vector must have a length no greater than unity since $\rho_{UA}(t)$ is positive semi-definite.

As we mentioned in the introduction, our numerical simulations indicate three distinct respects in which the evolution of $\rho_{UA}$ differs from 
the standard qubit evolution given in Eq.~(\ref{standardevol}).  
(i) The angular frequency of precession is reduced relative to the standard quantum mechanical value $\Omega$. 
(ii) There is a long-term dephasing of the state.  For the initial state we focus on here, with the spin in the $x$ direction, this dephasing
ultimately yields the completely mixed state.  
(iii) The length of the Bloch vector, which would normally maintain a constant value of unity, instead oscillates as the spin precesses, achieving its smallest value whenever 
the spin is directed along the $y$ axis.  The length of the Bloch vector indicates the purity of the 
state; so we see the state becoming mixed and then regaining its purity every half-cycle (long before the purity is reduced
permanently because of the dephasing).  Evidently what is happening is that
some of the information in the state is being shared temporarily with the environment and then returned to the $UA$ system.
When we trace over the environment to get $\rho_{UA}$, this shared component becomes invisible.  We call this shared
component of the state ``the ghost part.'' 

In the following three subsections we present our numerical results for each of these effects.

\subsection{Reduced precession frequency}

\begin{figure}[h]
\begin{center}
\includegraphics[scale = 1]{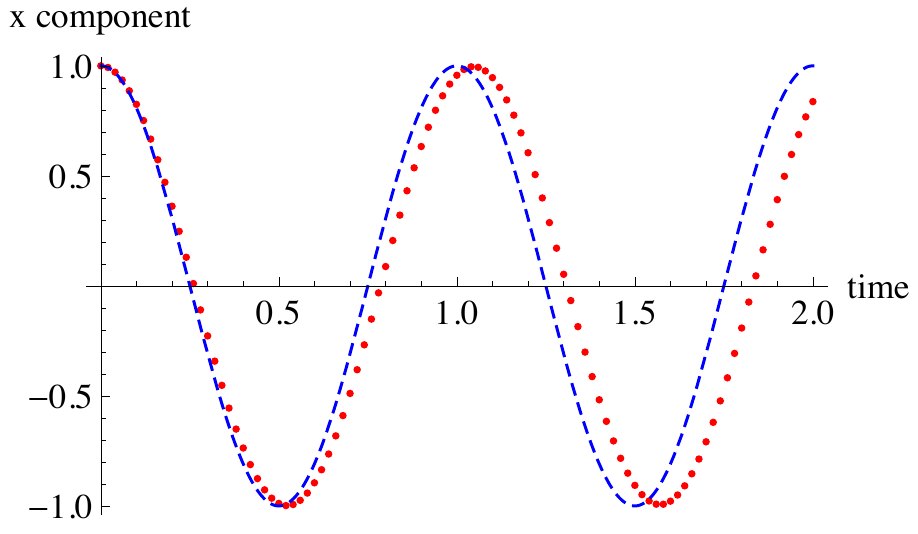}
\end{center}
\vspace{-3mm}
\caption{(Color online) The $x$ component of the Bloch vector as a function of time, compared to the standard quantum mechanical
prediction, for a spin initially in the $x$ direction and precessing around the $z$ axis.  The dots (red) show the numerical results,
and the dashed curve (blue) shows the standard quantum mechanical result.  In this plot $N=200$, $s=30$ and $\omega = 100$---we have chosen $\lambda = s/\omega$ to be relatively large to make
the effect visible---and we have set $\Omega = 2\pi$ so that time is measured in periods of the regular quantum mechanical precession.}
\label{frequencynofitfigure}
\end{figure}

Perhaps the most obvious way in which the precessing spin in our model departs from its behavior in standard quantum theory is that the
frequency of precession is reduced.
Numerically we find that the angular frequency depends on the value of our parameter $\lambda = {s}/\omega$, achieving the standard quantum mechanical
value $\Omega$ only as $\lambda$ goes to zero.  We show an example of the reduced frequency in Fig.~\ref{frequencynofitfigure},
which plots the $x$-component of the Bloch vector, $b_x$, as a function of time for a rather large value of $\lambda$.

\subsection{Long-term decoherence}

\begin{figure}[h]
\begin{center}
\includegraphics[scale = 1]{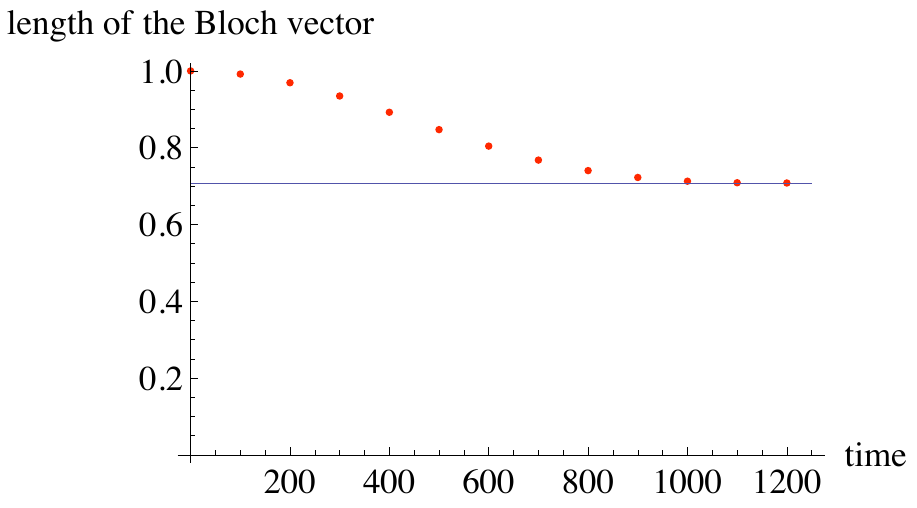}
\end{center}
\vspace{-3mm}
\caption{(Color online) For a spin initially directed at $45^\circ$ between the $x$ and $z$ axes and precessing around the
$z$ axis, the length of the Bloch vector
decays to $1/\sqrt{2}$ as the vector spirals into the $z$ axis.  The dots (red) show the numerical results, and the line (blue) 
is at the level $1/\sqrt{2}$.  Here $N=1400$, $s = 30$, $\omega = 300$, and $\Omega = 2\pi$, so that time is measured in 
precession periods.}
\label{decaynofitfigure}
\end{figure}

Over a sufficiently long time period, the precessing spin in our simulations decays to a stationary mixed state.
When the initial state is given by Eq.~(\ref{wholeinitialstate}), that is, when the spin is initially in the positive $x$ direction, and when
the precession is around the $z$ axis,
the Bloch vector eventually spirals into that axis, so that the final state of the $UA$ system is the completely mixed
state $(1/4)I_U \otimes I_A$.  If instead we take the initial state to have a non-zero value of $b_z$, we find that
in the evolving state $\rho_{UA}(t)$ the value of $b_z$ remains constant, but the $x$ and $y$ components of $\vec{b}$
again spiral into the $z$ axis, so that the $UA$ system finally settles into the constant state $(1/4)(I_U \otimes I_A + b_z I_U \otimes Z_A)$.  
That is, for any value of $b_z$ the phase coherence is eventually lost.  We find numerically that the decay time
depends on the environment dimension $N$, increasing with increasing $N$ before finally approaching a constant value when $N$ is large.  
We present an example in Fig.~\ref{decaynofitfigure}, for which the 
initial Bloch vector is $\vec{b} = (1/\sqrt{2}, 0, 1/\sqrt{2})$.  We plot there the length $b = \sqrt{b_x^2 + b_y^2 + b_z^2}$ of the Bloch vector, which seems to approach
the value $1/\sqrt{2}$, consistent with the picture of the vector spiraling into the $z$ axis.  In making the figure, we chose a value of $N$ that shows the large-environment limit.

\subsection{The ghost part}

\begin{figure}[h]
\begin{center}
\includegraphics[scale = 1]{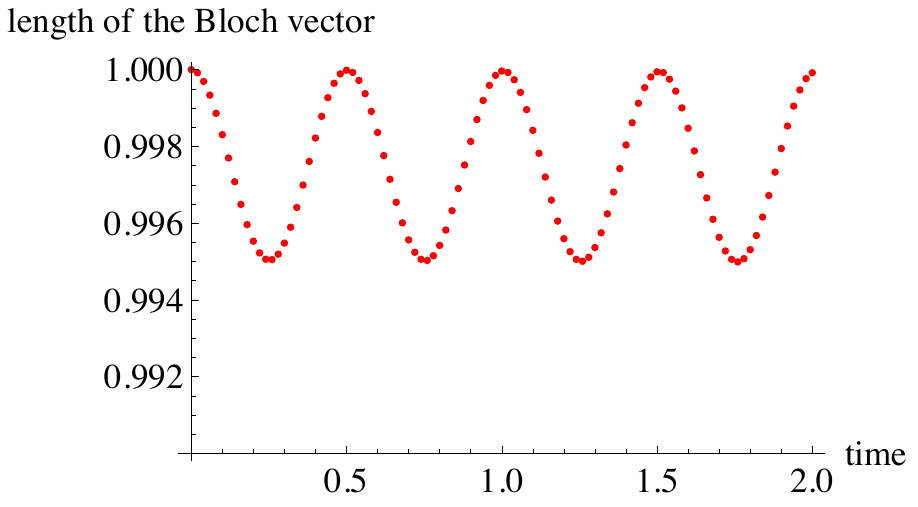}
\end{center}
\vspace{-3mm}
\caption{(Color online) The length of the Bloch vector varies periodically with a frequency equal to twice the precession frequency.  The dots (red) 
show the numerical results when the spin is initially in the positive $x$ direction.  Here $N=200$,
$s=30$, $\omega = 300$, and $\Omega = 2 \pi$, so that time is measured in precession periods.  The minima correspond to the times when the Bloch vector points in either
the positive or negative $y$ direction.  Notice that with these values of the parameters, the effect is small, with a reduction 
in length of about half a percent.}
\label{ghostnofitfigure}
\end{figure}

There is also a periodic change in the length of the Bloch vector as a function of time.  Our numerical results for $b$ as a
function of time
are shown in Fig.~\ref{ghostnofitfigure}, in which the spin is initially in the positive $x$ direction, and the time axis is in units of the usual quantum mechanical precession period $2\pi/\Omega$.  
Notice that the length achieves its minimum value twice in each cycle, corresponding to the times when the spin is pointing in the
positive or negative $y$ direction.  Indeed we find that $b_y$ never attains the value 1, while $b_z$ remains zero as 
expected.  It is as if the Bloch sphere were somewhat flattened along the $y$ axis, so that the Bloch vector precesses along the equator of 
the resulting oblate ellipsoid.  

The shortened Bloch vector associated with the $y$ axis indicates an increased entropy: some information has been lost.  But it has been
lost only temporarily, as it comes back in the next quarter-cycle.  Again, it appears that some kind of correlation has been temporarily set up between
the $UA$ system and the environment.  While it may be unusual for a correlation to automatically reverse itself, one can certainly find 
instances of such reversal in standard physics.  While a light pulse is reflecting off a mirror, for example, it is temporarily correlated with electrons in the 
mirror's silver coating, but once the reflection is complete the correlation has been undone and the pulse's state is, in the ideal case, as pure
as it was before the reflection.  (The effect we are seeing is similar to the phenomenon of ``false decoherence'' as described in Refs.~\cite{Leggett1, Leggett2, Anglin, Unruh}.)

One might wonder whether under some circumstances the correlation between the $UA$ system and the environment might become so thoroughly mixed up within the environment 
that it could never be undone, in which case the entropy of the $UA$ system would have increased permanently.  This seems to be
what happens after many cycles of precession---we eventually get the decoherence observed in Subsection IV.C.  But 
can the information get lost even when the spin is not precessing?  To investigate this question,
we have run simulations in which, after a quarter-cycle of precession, we turned off the local Stueckelbergian $S_{UA}$ and allowed 
the ubit and environment to continue to evolve according to $S_{EU}$ for thousands of precession periods while the $UA$ system remained 
``frozen'' along the $y$ axis.  We then turned 
$S_{UA}$ back on and let the system evolve for another quarter-cycle to see whether the Bloch vector would regain its full length.  
Indeed it did---the information had not been permanently lost.  Similar numerical experiments, in which the precession
axis was changed for the final part of the evolution, yielded similar results.  Now, it is certainly possible to cause the
correlation to be lost in the environment while the spin is not precessing.  
It is sufficient to change $B_{EU}$ itself in the middle of the numerical run.  
But normally we can understand a time-dependent Hamiltonian as arising from a stationary Hamiltonian acting on a larger system.  (We
include in the system whatever is causing the Hamiltonian to change.)  So it does not seem unreasonable to assume
a time-independent operator $B_{EU}$ as we do here, in which case it seems that the ``ghost part'' can be made to return to the
$UA$ system, at least in the short term.     

\subsection{A second special axis}

In the example of a precessing spin, it is clear that the precession axis, that is, the axis along which the magnetic field vector lies, plays a
special role.  In the ubit model, it turns out that there is a second special axis: it is the axis that in the complex theory is associated with
the purely imaginary Pauli matrix.  (The standard convention, which we are using, is to call this axis the $y$ axis.)  The specialness of
this second axis has not been evident in the numerical experiments described above, because in those experiments the precession axis has always been the $z$ axis, whose Pauli matrix $Z$ is purely real.
In fact the results of those experiments would be essentially unchanged if we were to choose any precession axis in the $xz$ plane,
since any real linear combination of $X$ and $Z$ is also real.  
However, when the 
precession axis is the $y$ axis, {\em all} of the above effects disappear.  There is neither a periodic nor a long-term change in the length
of the Bloch vector, and there is no reduction in the frequency of precession relative to standard quantum mechanics.  
Moreover, for a precession
axis 
intermediate between the 
$xz$ plane and the $y$ axis, the above effects all appear but are not as large as when the precession axis is in the $xz$ plane.
Consider the decoherence, for example.  For an intermediate precession axis, we observe that
the Bloch vector again spirals into the axis of precession without changing its component along that axis. 
That is, we observe what in the complex theory would be called a loss of phase coherence in the energy basis.  But the coherence
time becomes longer as the precession axis becomes more parallel to the $y$ axis.  In this way the $y$ axis plays a special role.

There is, in addition, one effect pertaining to the $y$ axis that has nothing to do with precession.  Suppose we have no magnetic field---that is, we set $S_{UA}$ to zero---and we choose the initial state $\rho_{UA}(0)$ to be 
$(1/4)(I_U \otimes I_A + J_U \otimes J_A)$.  That is, we try to start the spin in the positive $y$ direction.  Then one finds that over an extremely short time,
the Bloch vector shrinks to a shorter length (still in the same direction).  This is what one expects from Section III: the coefficient of $J_U$ 
quickly decreases by the factor $1 - \lambda^2$.  A Bloch vector of this length is in fact even shorter than what we would get
by starting the spin in the $x$ direction and letting it precess for a quarter cycle.  As we will see in the next section,
the length in the latter case is $1- \lambda^2/2$.  In either case, a literal reading of the ubit model would seem to say that it is impossible to prepare a pure state of spin in the $y$ direction.  Instead, we can prepare only
a mixed state in that direction.  If we accept this reading, the Bloch sphere really is
flattened into an oblate ellipsoid.  States that lie beyond the boundary of this ellipsoid are simply inaccessible.  In Section VI we will introduce
an alternative interpretation in which a pure state in the $y$ direction is possible, but even in this reinterpretation the predicted physics depends
on which axis in space we associate with the imaginary Pauli matrix $Y$.   

Evidently
in the ubit model, in order to completely describe the dynamical situation of a spin-1/2 particle, one needs to specify not only
the direction and strength of the magnetic field, but also the direction of the ``$y$ axis,'' which now becomes physically important.  
Conceivably the experimenter would have
control over this second axis, just as she has control over the magnetic field.  Or possibly the second axis would be beyond the experimenter's
independent control; for example, a law of nature could force a relationship between this second axis and the magnetic field axis.  

In the case of spin precession, the second special axis is an axis in space. 
But for other physical realizations of a qubit, e.g. a two-level atom, the ``direction'' one associates with the imaginary Pauli matrix $Y$
is not a direction in space; usually it is associated with a particular equal-magnitude superposition of the ground and excited states.  
Moreover, we normally have a unitary symmetry that allows us to freely re-express any problem in whatever basis we choose---the choice of basis
has no physical significance.  However, the ubit model forces us to treat separately the real and imaginary parts of a Hamiltonian or a density matrix, and this separation between
real and imaginary parts could be changed simply by changing the basis.  To put it in other words, for any given Hamiltonian there are many distinct Stueckelbergians, depending on the basis in which the Hamiltonian is written.  
Section VII explores this question further in the case of higher dimensions.  For now, though, we try to explain analytically the three effects
described in the preceding subsections.  

\section{A Precessing Spin---Analytical Treatment}

We begin our analysis with Eqs.~(\ref{projrho}) and (\ref{pseudoH}), in which we have already assumed that the state $\hat{\rho}_\parallel$ is continually being projected onto
the space of matrices that commute with $S_{EU}$.  We write those equations again here:
\begin{equation}
\hat{\rho}_\parallel(t) = \frac{1}{2N}\sum_j |\Psi_j\rangle\langle\Psi_j| \otimes \sigma_j(t);
\end{equation}
\begin{equation}  \label{Aeq}
\frac{d\sigma_j}{dt} = -i[\nu_jK_A + iL_A, \sigma_j].
\end{equation}
For the particular case we are considering now, the local Stueckelbergian is
\begin{equation}
S_{UA} = I_U \otimes L_A - J_U \otimes K_A = -\frac{\Omega}{2}J_U \otimes Z_A,
\end{equation}
so that $K_A = (\Omega/2)Z_A$ and $L_A = 0$.
Inserting this expression into Eq.~(\ref{Aeq}) gives us
\begin{equation}  \label{specialA}
\frac{d\sigma_j}{dt} =-\frac{i\nu_j\Omega}{2} (Z \sigma_j - \sigma_j Z),
\end{equation}
where again $\nu_j =- i \langle \Psi_j | I_E \otimes J_U | \Psi_j \rangle$. We can solve Eq.~(\ref{specialA})
to get
\begin{equation}
\sigma_j(t) = e^{-i(\nu_j\Omega/2)Zt} \sigma_j(0) e^{i(\nu_j\Omega/2)Zt}.
\end{equation}
We are assuming an initial state given by Eq.~(\ref{wholeinitialstate}), in which each $\sigma_j(0)$ is equal to $(1/2)(I_A + X_A)$.
In that case we have
\begin{equation}
\sigma_j(t) = \frac{1}{2}\left[I_A + \cos(\nu_j\Omega t)X_A + \sin(\nu_j\Omega t)Y_A\right],
\end{equation}
where $Y_A = iJ_A$ is the usual imaginary Pauli matrix.  (Again $\sigma_j$ can have a nonzero imaginary part.  But in $\hat{\rho}_\parallel$ all
imaginary contributions will cancel.)  
The effective density matrix of the whole system is 
\begin{equation}  \label{startingpoint}
\hat{\rho}_\parallel(t) =  \frac{1}{4N}\sum_j |\Psi_j\rangle\langle\Psi_j| \otimes \left[I_A + \cos(\nu_j \Omega t)X_A + \sin(\nu_j\Omega t)Y_A\right].
\end{equation}

Our strategy will be to try to isolate each of the three effects described above by considering different 
terms of the perturbation expansion of Eq.~(\ref{startingpoint}).   (i) To see the frequency reduction, we expand $\nu_j$ to second order while
approximating $|\Psi_j\rangle\langle\Psi_j|$ in Eq.~(\ref{startingpoint}) with its unperturbed value.  (ii) A spread in the values of $\nu_j$ would lead to interference when we do the sum in Eq.~(\ref{startingpoint}), which would appear as decoherence.  But the $\nu_j$'s begin to diverge from each other only in third order.   Therefore, to isolate the long-term decoherence, we expand $\nu_j$ to third order while continuing to
treat $|\Psi_j\rangle\langle\Psi_j|$ as unperturbed.  (iii) The ghost part represents a correlation between the ubit and the environment.  So to see the ghost part, we will expand $|\Psi_j\rangle\langle\Psi_j|$ in Eq.~(\ref{startingpoint}) out to lowest nontrivial order (it will be first order)
while restricting our approximation for $\nu_j$ to second order so as to avoid the complications of decoherence.

\subsection{Reduced precession frequency}

\begin{figure}[h]
\begin{center}
\includegraphics[scale = 1]{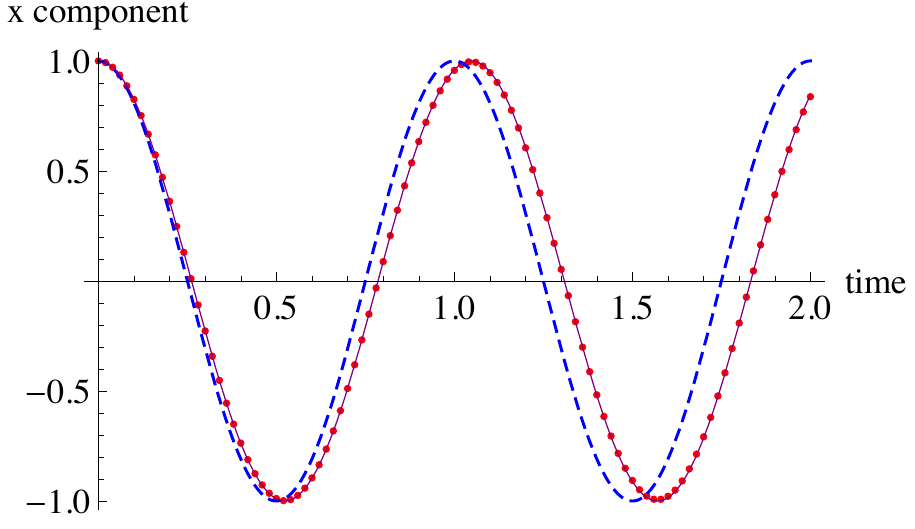}
\end{center}
\vspace{-3mm}
\caption{(Color online) Comparison of the numerical data with our analytic result for the $x$ component of the Bloch vector, when the spin is initially
in the $x$ direction and is precessing around the $z$ axis.  The dots (red) are the numerical results and the continuous curve (purple)
shows our analytic prediction. Again the dashed curve (blue) represents the standard quantum
mechanical precession. Here $N = 200$, $s=30$, $\omega = 100$, and $\Omega = 2 \pi$, so that time is measured in precession periods.}
\label{frequencyfigure}
\end{figure}

We begin with Eq.~(\ref{nudef}) for $\nu_j$ and expand each $|\Psi_j^+\rangle$ in that equation out to second order in $\lambda = {s}/\omega$.  Starting with the
expansion of $|\Psi_j\rangle$ given in Appendix A, we find that 
\begin{equation}
\nu_j = \nu_n^\pm = -i \langle \Psi_n^\pm | I_E \otimes J_U | \Psi_n^\pm \rangle = \pm \left( 1 - \frac{\lambda^2}{2}\sum_k \left| \left\langle \Phi_n^+ \left| V \right| \Phi_k^-\right\rangle \right|^2  \right),
\end{equation}
where we have replaced the single index $j$ with the pair of indices $n$ and $\pm$.  The $+$ and $-$ refer to the subspaces
in which the operator $G$ takes positive and negative
values, respectively.  
As we have done before, for each value of $n$ and $k$ we replace $\left| \left\langle \Phi_n^+ \left| V \right| \Phi_k^-\right\rangle \right|^2$ with its ensemble average,
$1/N$, thereby arriving at
\begin{equation}
\nu_n^\pm = \pm \left[ 1 - \frac{\lambda^2}{2} \right] = \nu^\pm.
\end{equation}
At this order of perturbation theory there is no dependence on $n$.  The factor $|\Psi_j\rangle
\langle \Psi_j |$ in Eq.~(\ref{startingpoint}) we treat as unperturbed, so that $\sum |\Psi_n^\pm\rangle \langle \Psi_n^\pm|$ can be
replaced with $\sum |\Phi_n^\pm\rangle \langle \Phi_n^\pm| = P_\pm = (1/2)i I_E \otimes (I_U \mp J_U)$.  With these substitutions, one finds that 
\begin{equation}
\hat{\rho}_\parallel(t) = \frac{1}{4N}\left\{ I_E I_U I_A + \cos\left[\left(1 - \frac{\lambda^2}{2} \right) \Omega t\right]
I_E I_U X_A + \sin\left[\left(1 - \frac{\lambda^2}{2} \right) \Omega t\right]I_E J_U J_A \right\},
\end{equation}
where we have left out the tensor product symbols.  Tracing over the environment, we get that the density matrix of the $UA$ 
system is
\begin{equation}
\rho_{UA}(t) = \frac{1}{4}\left\{ I_U I_A + \cos\left[\left(1 - \frac{\lambda^2}{2} \right) \Omega t\right]
 I_U X_A + \sin\left[\left(1 - \frac{\lambda^2}{2} \right) \Omega t\right] J_U J_A \right\}.
\end{equation}
That is, with these approximations the spin precesses as usual but with its frequency reduced by the factor 
$1 - \lambda^2/2$.  In Fig.~\ref{frequencyfigure} we compare this theoretical prediction with the numerically observed evolution.  

\subsection{Long-term decoherence}

\begin{figure}[h]
\begin{center}
\includegraphics[scale = 1]{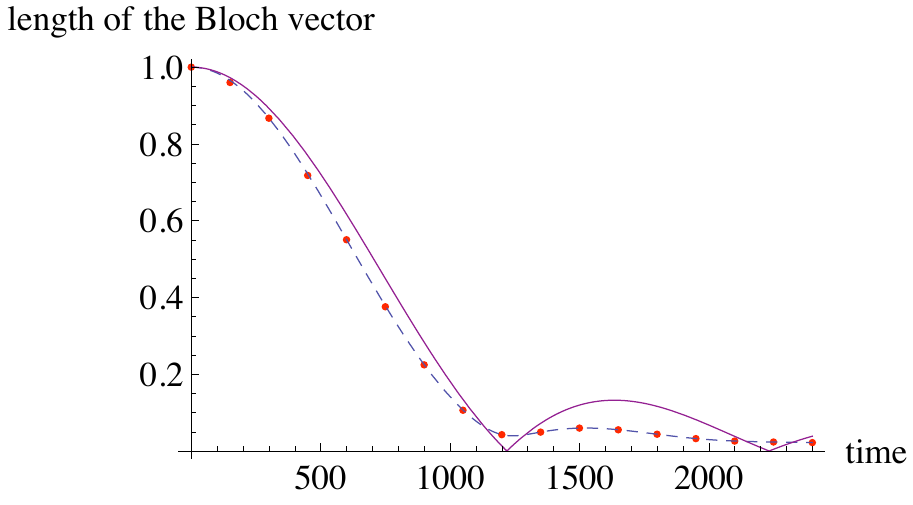}
\end{center}
\vspace{-3mm}
\caption{(Color online) The spin begins in the $x$ direction and spirals into the $z$ axis as it precesses around that axis.  Here we plot the length
of the Bloch vector as a function of time.  The dots (red) show the numerical results, and the dashed curve (blue) 
gives the prediction of Eq.~(\ref{rhodeco}) with the $\nu_j$'s computed numerically.   The solid curve (purple)
shows our analytic approximation given by Eqs.~(\ref{rhodeco2}) and (\ref{eff}), in which we have made a 
simple approximation for the values of $\nu_j$.  Here $N=1400$, $s=30$, 
$\omega = 300$, and $\Omega = 2 \pi$, so that time is measured in precession periods.  The ``bounce'' we see here does not show up in Fig.~\ref{decaynofitfigure}, because
in that case a small wobble around the $z$ axis does not significantly affect the Bloch vector's length.}
\label{decayfigure}
\end{figure}

In Eq.~(\ref{startingpoint}), both the $EU$ factor and the $A$ factor depend on $j$, so that we cannot in general separate these two 
parts of the system when we do the sum.  However, to try to get an analytic handle on the decoherence, 
we assume that within each of the two main subspaces the $EU$ factor is not significantly correlated with the $A$ factor; so within
each subspace we can say that the average of the product
is the product of the averages.  If we also continue to assume that we can replace each $|\Psi_j\rangle\langle \Psi_j|$ with 
its unperturbed value $|\Phi_j\rangle\langle \Phi_j|$, we get
\begin{equation} \label{longeq}
\begin{split}
\hat{\rho}_\parallel(t) = \frac{1}{4N^2} & \left\{ P_+ \otimes \sum_n \left[ I_A + \cos(\nu_n^+ \Omega t)X_A + \sin(\nu_n^+ \Omega t)Y_A \right] \right.\\
&+\left. P_- \otimes \sum_m \left[ I_A + \cos(\nu_m^- \Omega t)X_A + \sin(\nu_m^- \Omega t)Y_A \right] \right\}. 
\end{split}
\end{equation}
Again using $P_\pm = (1/2)iI_E \otimes (I_U \mp J_U)$, we can rewrite this expression as
\begin{equation}  \label{rhodeco}
\hat{\rho}_\parallel(t) = \frac{1}{4N^2} \sum_n \left[I_EI_UI_A + \cos(\nu_n^+ \Omega t)I_EI_UX_A + \sin(\nu_n^+ \Omega t)I_EJ_UJ_A \right].
\end{equation}
In writing this last equation we have used the fact, mentioned earlier, that the vectors $|\Psi_j\rangle$ come in complex-conjugate pairs,
and that the real values $\nu_j$ given by Eq.~(\ref{nudef})
come in pairs with equal magnitudes and opposite signs.  This fact is what allows us to combine the two sums in Eq.~(\ref{longeq}) into a single sum.  Numerical tests confirm that Eq.~(\ref{rhodeco}) yields a very close approximation to $\rho_{UA}$.  For example, 
in Fig.~\ref{decayfigure} the dashed curve shows the length of the Bloch vector as predicted by Eq.~(\ref{rhodeco})---with the values
of $\nu_j$ determined numerically---while the dots represent the numerical
results obtained directly from Eq.~(\ref{exactrhoUA}).  The good 
agreement provides evidence in support of our assumption of continual projection,
made in Subsection III.D,
as well as for the specific assumptions leading to Eq.~(\ref{rhodeco}) in the present section.  Still, we would prefer
an equation that does not require an exact determination of $\nu_n^+$.  So we take our approximation a step further.  

In order to evaluate $\nu_n^+$ to third order, we find it convenient to use the relation
\begin{equation}
-i\langle \Psi_n |I_E \otimes J_U |\Psi_n \rangle = g_n - \lambda \langle \Psi_n | V | \Psi_n\rangle,
\end{equation}
which comes from Eqs.~(\ref{one}) and (\ref{two}).
Thus it is sufficient to expand $g_n$ to third order and $|\Psi_n\rangle$ to second order.  
On carrying out this expansion, we find that the third-order contribution to $-i\langle\Psi_n^+| I_E \otimes J_U |\Psi_n^+\rangle$
is
\begin{equation} \label{thirdorder}
\begin{split}
-i\langle\Psi_n^+| I_E \otimes J_U |\Psi_n^+\rangle^{(3)} =  \frac{\lambda^3}{2}\Bigg[ &\left\langle \Phi_n^+\right| V \left| \Phi_n^+\right\rangle
\left\langle \Phi_n^+\right| VP_-V \left| \Phi_n^+\right\rangle
- \left\langle \Phi_n^+\right| VP_-VP_-V \left| \Phi_n^+\right\rangle  \\
& - \sum_k \frac{\left| \left\langle \Phi_n^+\right| VP_-V \left| \Phi_k^+\right\rangle  \right|^2}{\left\langle \Phi_n^+\right| V \left| \Phi_n^+\right\rangle-\left\langle \Phi_k^+\right| V \left| \Phi_k^+\right\rangle} \Bigg].
\end{split}
\end{equation}

We consider the three terms in this expression separately.  First,
following the reasoning
in Eq.~(\ref{earlyapprox}) we approximate $\left\langle \Phi_n^+\right| VP_-V \left| \Phi_n^+\right\rangle$ as $1$, so that
the first term square brackets in Eq.~(\ref{thirdorder}) can be approximated as $\left\langle \Phi_n^+\right| V \left| \Phi_n^+\right\rangle$, 
which we have called $v_n$.  We can write the second term as
\begin{equation}
 \left\langle \Phi_n^+\right| VP_-VP_-V \left| \Phi_n^+\right\rangle = \sum_k \left| \left\langle \Phi_n^+\right| V \left| \Phi_k^-\right\rangle \right|^2
 \left\langle \Phi_k^-\right| V \left| \Phi_k^-\right\rangle.
 \end{equation}
 Again we use $\left| \left\langle \Phi_n^+\right| V \left| \Phi_k^-\right\rangle \right|^2 \approx 1/N$, so that we are left with a 
 sum over the eigenvalues of the negative-subspace part of $V$.  Those eigenvalues have a typical size that does not depend on $N$,
 but their ensemble average is zero, and we expect their sum to be of order $\sqrt{N}$ because of random
 fluctuations.  Thus the whole term diminishes as $1/\sqrt{N}$, 
 and since we assume a large environment dimension we take this term to be zero.  
 
 The third term in Eq.~(\ref{thirdorder}) is more complicated.  We can approximate the numerator as $1/N$.  The denominator, which
 we can write as $v_n - v_k$, can be small, so that the sum might depend crucially on the spacing of the values $v_k$.  Those values follow a semicircle
 distribution, but this fact does not tell us how the difference $v_n - v_k$ is distributed.  To get a somewhat crude approximation, we 
 ignore the issue of the spacing of values and simply replace the sum with an integral, assuming a semicircle distribution, and take the Cauchy principal value of the integral.  This gives us
 \begin{equation}
 \sum_k \frac{1/N}{v_n - v_k} \approx \int_{-2}^{2}
 \frac{1}{N}\,\frac{1}{v_n - v}\eta(v) dv = \frac{v_n}{2}
 \end{equation}
 for the semicircle distribution $\eta(v) = (N/\pi)\sqrt{1 - (v/2)^2}$.  The third term then combines 
 quite simply with the first term to give us
 \begin{equation}
- i\langle\Psi_n^+| I_E \otimes J_U |\Psi_n^+\rangle^{(3)} \approx \frac{\lambda^3 v_n}{4}.
 \end{equation}
 
 We can now put this result back into Eq.~(\ref{rhodeco}) and convert the sum to an integral, again assuming a semicircle distribution
 for $v_n$, obtaining a result for the decay similar to what we saw in Section III.
 \begin{equation} \label{rhodeco2}
 \rho_{UA} \approx \frac{1}{4} \left\{I_UI_A + f(t)\left[ \cos\left(\xi\Omega t\right)I_UX_A + \sin\left(\xi\Omega t\right)J_UJ_A \right] \right\},
 \end{equation}
 where
 \begin{equation} \label{eff}
 f(t) =\left| \frac{2 J_1\left(\lambda^3 \Omega t/2\right)}{\lambda^3 \Omega t/2} \right|
 \end{equation}
and $\xi = 1 - \lambda^2/2$ is the frequency reduction factor we computed in the preceding subsection.  
Fig.~\ref{decayfigure} shows the length of the Bloch vector as a function of time, as computed from Eq.~(\ref{rhodeco2}), and compares this approximation with
the numerical values and with Eq.~(\ref{rhodeco}).  Clearly our approximation of the $\nu_j$'s is not ideal, but it seems to give us at least a reasonable estimate of the 
coherence time $\tau$,
which is of order 
\begin{equation}
\tau \approx \frac{1}{\lambda^3\Omega}.
\end{equation}
(In fact $\tau$ has to be over five times this value to make $f(\tau)$ less than $1/e$, though of course the curve is not
exponential.)  
Moreover, the detailed shape of the curve traced out by our numerical results in Fig.~\ref{decayfigure} surely depends on the specific
model of the environment we have chosen, whereas the scaling of the coherence time with $\lambda$ and $\Omega$ has a better chance of carrying over to other
models.   

\subsection{The ghost part}

To understand the ghost part, we begin by replacing each $\nu_j$ in Eq.~(\ref{startingpoint}) with its second-order value, which is
 $\pm\xi$ for $\nu_n^\pm$.  (Again $\xi = 1 - \lambda^2/2$.)  This approximation
allows us to write $\hat{\rho}_\parallel(t)$ as
\begin{equation}  \label{rhorho}
\hat{\rho}_\parallel(t) = \frac{1}{2N}\left(\sum_n \left| \Psi_n^+ \right\rangle \left\langle \Psi_n^+\right| \otimes \sigma^+(t)
+ \sum_n \left| \Psi_n^- \right\rangle \left\langle \Psi_n^-\right| \otimes \sigma^-(t) \right),
\end{equation}
where
\begin{equation}
\sigma^\pm = \frac{1}{2}\left[ I_A + \cos(\xi\Omega t) X_A \pm \sin(\xi\Omega t) Y_A \right].
\end{equation}
Expanding $\left| \Psi_n^+\right\rangle$ and $\left| \Psi_n^-\right\rangle$ to first order in $\lambda$, we find
\begin{equation}
\sum_n \left| \Psi_n^+ \right\rangle \left\langle \Psi_n^+\right|  = P_+ + \frac{\lambda}{2}\left[P_- V P_+ + P_+ V P_- \right]
\end{equation}
and
\begin{equation}
\sum_n \left| \Psi_n^- \right\rangle \left\langle \Psi_n^-\right|  = P_- - \frac{\lambda}{2}\left[P_- V P_+ + P_+ V P_- \right].
\end{equation}
Upon inserting these expressions in Eq.~(\ref{rhorho}) we get
\begin{equation}  \label{ghostrho}
\hat{\rho}_\parallel(t) = \frac{1}{4N}\left[ I_E  I_U  I_A + \cos(\xi\Omega t) I_E I_U  X_A
+ \sin(\xi\Omega t)\left\{ I_E J_U  + \lambda \left( P_- B P_+ + P_+ B P_- \right)\right\}  J_A  \right],
\end{equation}
where we have again left out the tensor product symbols.

\begin{figure}[h]
\begin{center}
\includegraphics[scale = 1]{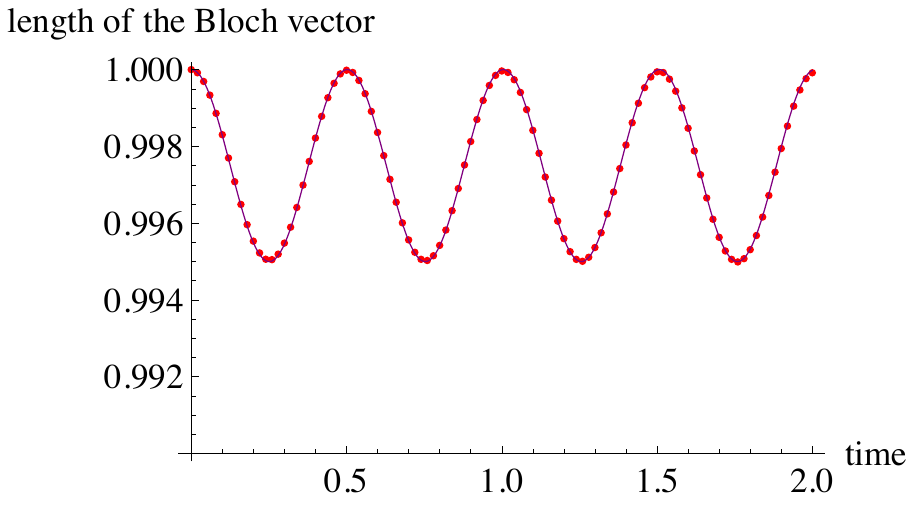}
\end{center}
\vspace{-3mm}
\caption{(Color online) Comparison of our analytic result with the numerical data for the periodic variation of the 
length of the Bloch vector.  The spin starts in the positive $x$ direction and precesses around the $z$ axis.
The dots (red) show the numerical results and the curve (purple) is obtained from Eq.~(\ref{ghostrhoUA}).  
Again the minima occur when the Bloch vector is in the positive or negative $y$ direction.  Here
$N=200$, $s=30$, $\omega = 300$, and $\Omega = 2 \pi$, so that time is measured in precession periods.}
\label{ghosttempfigure}
\end{figure}

The term proportional to $\lambda$ is what we are calling the ghost part.  Notice that it accompanies what we would normally
think of as the $y$ component of the Bloch vector (that is, the part proportional to $J_U \otimes J_A$).  Except for the ghost part
and the frequency reduction,
the above expression is identical to the standard quantum mechanical evolution given in Eq.~(\ref{standardevol}).  Of course 
our expression for the ghost part is valid only to first order in $\lambda$.  If we expand each $|\Psi_j\rangle$ out to second order
and trace over the environment, we find that the density matrix of the $UA$ system is
\begin{equation}  \label{ghostrhoUA}
\rho_{UA}(t) = \frac{1}{4}\left[ I_U I_A + \cos(\xi\Omega t) I_U X_A + \left(1 - \frac{\lambda^2}{2}\right) \sin(\xi\Omega t) J_U J_A \right],
\end{equation}
which shows the shortening of the $y$ component of the Bloch vector.  Here we have assumed a large environment, so that
$\hbox{Tr}_E B_{EU}$ can be taken to be zero.  (The ensemble average of that partial trace is zero, with fluctuations of order unity.
The factor of $1/(4N)$ normalizing the density matrix renders such fluctuations negligible.)  Under this assumption, the ghost
part has disappeared.  Fig.~\ref{ghosttempfigure} compares Eq.~(\ref{ghostrhoUA}) with our numerical results.

 The picture that emerges, then, is that part of the $y$ component of the Bloch vector has been lost, but it has been replaced
 with the ghost part, which represents a correlation between the ubit and the environment.  When one traces over the environment,
 what remains for the $UA$ system is a Bloch sphere that has been flattened along the $y$ axis by the factor 
 $1 - \lambda^2/2$.  However, there is an alternative interpretation of the ghost part that we find more appealing; 
 this alternative interpretation is the subject of the next section.
 
 \section{The Modified-Ubit Interpretation}
 
 We have assumed that Alice can perform any measurement on the $UA$ system.  One such measurement
 for the case of a spin-1/2 particle is to test whether the spin is in the positive $y$ direction.  In ordinary quantum mechanics
 this test would be represented by the projection operator $(1/2)(I + Y)$.  The direct translation of this operator into 
 real-vector-space terms, according to 
 the prescription of Section II, is the rank-two projection operator $(1/2)(I_U \otimes I_A + J_U \otimes J_A)$.  
 If this measurement were performed on the completely mixed state and the ``yes'' outcome were obtained, the state of the $UA$ system would be
 collapsed into the state $\rho_{UA} = (1/4)(I_U \otimes I_A + J_U \otimes J_A)$, which is the real-vector-space version 
 of spin in the positive $y$ direction.  But this state cannot persist for any nonzero duration
 under the projection assumption of Subsection III.D.  So it would seem that Alice cannot prepare a pure state of
 spin in the $y$ direction, as we noted earlier.  
 
 But there is another way Alice might try preparing a spin state in the $y$ direction.  She could perform the measurement
 $(1/2)(I_U \otimes I_A + I_U \otimes X_U)$---testing for spin in the positive $x$ direction---and upon obtaining the outcome ``yes" she
 could allow the spin to precess around the $z$ axis for a quarter-cycle, thereby preparing the effective $EUA$ state
 \begin{equation}  \label{rhoy}
\hat{\rho}_y = \frac{1}{4N}\left[ I_E  I_U  I_A 
+\left\{ I_E J_U  + \lambda\left( P_- B P_+ + P_+ B P_- \right)\right\}  J_A  \right], 
\end{equation}
 in accordance with Eq.~(\ref{ghostrho}).  Moreover, if at some later time she wanted to test for this state, she could do so
 by first allowing the spin to precess by a quarter-cycle (in the same direction as before) and then performing the measurement
 $(1/2)(I_U \otimes I_A - I_U \otimes X_A)$, that is, a test corresponding to the negative $x$ direction.  (Recall our numerical
 experiments in which the ghost part could be recovered after a long time during which there was no precession.)  This sequence of 
 operations is perfectly permissible according to our rules; so it should count as a valid measurement.  In this sense the 
 state $\hat{\rho}_y$ given in Eq.~(\ref{rhoy}) acts like what we would normally think of as a pure state of spin in the $y$ direction.  
 One can test for this state and get the ``yes'' outcome with unit probability.  So it seems that Alice can prepare a pure spin state in 
 the $y$ direction after all.  It does not look like a pure state when one traces over the environment---it looks like a mixed state---but
 it acts like a pure state.  Again, this effect can be seen as an example of ``false decoherence" \cite{Leggett1, Leggett2, Anglin, Unruh},
 in which part of the environment adiabatically follows the evolution of the system of interest.  In cases of false decoherence it is misleading 
 simply to trace out
 the environment, and we seem to have the same kind of situation here.  
 
 In our alternative interpretation, then, the Bloch sphere is not flattened.  To first order in $\lambda$, a general pure state of spin would be
 expressed in this interpretation by the $EUA$ density matrix
 \begin{equation}  \label{altspin}
 \hat{\rho}_{\vec{b}} = \frac{1}{4N}\left[ I_E  I_U  I_A  + b_x I_E I_U X_A + b_y 
\left\{ I_E J_U  + \lambda\left( P_- B P_+ + P_+ B P_- \right)\right\}  J_A  + b_z I_E I_U Z_A \right], 
\end{equation}
for some unit vector $\vec{b} = (b_x, b_y, b_z)$.  (Eq.~(\ref{altspin}) is valid only in the special case we have been considering in which the environment
starts out in the completely mixed state.  For a more general initial state the form would be different, as we will see in the 
following paragraphs.)
Alice can prepare any such state, and if $|\vec{b}| = 1$ she can test for the state.  
It is only that the mathematical description of the state is not what we would have expected, since it implicates part of 
the environment.

We now spell out the alternative interpretation, which we call the ``modified-ubit interpretation,'' more completely and for a more general case.  Let $A$ be a $d$-dimensional system, and 
let us assume for now that $\nu_j$ can be approximated by its second-order expansion.  (This assumption will be
relaxed shortly.)  Eq.~(\ref{projrho}) gives the general form of a state consistent with our projection assumption.  Our observer
Alice has no direct control over the environment portion of the state, but according to our initial assumptions she can at least 
prepare a state $\sigma(0)$ of the system $A$.  If she does so, the effective state of the whole system will have the form  
\begin{equation}  \label{initialstate}
\hat{\rho}_\parallel(0) = \frac{1}{2}\sum_n  c_n \left(|\Psi_n^+\rangle\langle \Psi_n^+ | +   |\Psi_n^-\rangle\langle \Psi_n^- | \right) \otimes \sigma(0),
\end{equation}
where $\sigma(0)$ is a real, positive semi-definite matrix with unit trace, and the non-negative coefficients $c_n$ sum to unity.  
These coefficients are determined
by the initial state of the environment and ubit (the initial state of the environment is no longer assumed to be the completely mixed state), and we assume that Alice has had
no control over their values.  

Alice can manipulate the initial state (\ref{initialstate}) by choosing a Stueckelbergian $S_{UA}$ of the form
$S_{UA} = I_U \otimes L_A - J_U \otimes K_A$.  In general the application of this Stueckelbergian would cause 
the matrix $\sigma(0)$ to evolve in a different way for each term $|\Psi_n^\pm\rangle\langle\Psi_n^\pm|$, but as we have seen,
to second order in $\lambda$ there are only two distinct value of $\nu_j$, namely, $\nu^\pm = \pm \xi$.  So the initial
state evolves into 
\begin{equation}  \label{evolvedstate}
\hat{\rho}_\parallel(t) = \frac{1}{2}\sum_n c_n \left(|\Psi_n^+\rangle\langle \Psi_n^+ | \otimes \sigma(t) +   |\Psi_n^-\rangle\langle \Psi_n^- |\otimes \overline{\sigma(t)} \right),
\end{equation}
where
\begin{equation}  \label{sigmat}
\sigma(t) = e^{-iH't} \sigma(0) e^{iH't}
\end{equation}
and the effective Hamiltonian is $H' = \xi K_A + i L_A$ in accordance with Eq.~(\ref{pseudoH}). 

We can rewrite Eq.~(\ref{evolvedstate}) as
\begin{equation}  \label{gammaequation}
\hat{\rho}_\parallel = \frac{1}{2}\left( \Gamma^+ \otimes \sigma + \Gamma^- \otimes \overline{\sigma} \right),
\end{equation}
where 
\begin{equation}
\Gamma^\pm = \sum_n c_n |\Psi_n^\pm\rangle\langle \Psi_n^\pm |
\end{equation}
and we have written $\sigma(t)$ simply as $\sigma$.  
Now, every complex $d \times d$ positive semi-definite matrix with unit trace can be written in the form given in the right-hand
side of Eq.~(\ref{sigmat}) for some real $\sigma(0)$
and some Hermitian $H'$.  
Because Alice can control $\sigma(0)$ and $H'$ (and $t$), she can determine the matrix $\sigma$ in 
the state (\ref{gammaequation}).  So we can think of this state as the result of Alice's preparation.  She determines
$\sigma$, but she does not control $\Gamma^\pm$ which is determined by the environment.  
  
We can also write Eq.~(\ref{gammaequation})
in the following way:
\begin{equation}   \label{newsplit}
\hat{\rho}_\parallel = \frac{1}{2} \left( {\mathcal I} \otimes  \hbox{Re}\,\sigma + {\mathcal J} \otimes \hbox{Im}\,\sigma  \right),
\end{equation}
where
\begin{equation}
{\mathcal I} =  \Gamma^+ + \Gamma^-  \hspace{1cm} \hbox{and} \hspace{1cm} 
{\mathcal J} =i \left( \Gamma^+ - \Gamma^- \right).
\end{equation}
The matrix $\sigma$ evolves according to the equation
\begin{equation}  \label{newSchrodinger}
\frac{d\sigma}{dt} = [ -iH', \sigma].
\end{equation}
In this respect $\sigma$ behaves like a density matrix.  Eq.~(\ref{newsplit}) is reminiscent of 
Eq.~(\ref{complextoreal}) in Section II, but 
the matrices ${\mathcal I}$ and ${\mathcal J}$ act on the whole $EU$ space rather than just on $U$. 

We now want to identify, in effect, a ``modified ubit'' $U'$---it will involve the environment---in terms of which the 
operators ${\mathcal I}$ and ${\mathcal J}$ will appear as tensor products.  The modification is expressed by an orthogonal transformation $O$ acting on the $EU$ system.  Our idea is that the application of $O$, followed by a trace over the environment, should leave us with the
state of $U'$ (or of $U'A$ if the system $A$ was included initially).  
In order that ${\mathcal I}$ and ${\mathcal J}$ be turned into tensor products, we want $O$ to have the following effect:
\begin{equation}  \label{thisequation}
O |\Psi_n^+\rangle\langle \Psi_n^+ | O^T = \frac{1}{2}|n\rangle\langle n| \otimes (I_U - i J_U),
\end{equation}
where the vectors $|n\rangle$ constitute an orthonormal (real) basis for the environment.  (For our purposes it does not 
matter which basis we choose.)  We construct such a transformation in Appendix C.  

From Eq.~(\ref{thisequation}) it follows that 
\begin{equation}   \label{IJtransform}
O {\mathcal I} O^T = \rho_E \otimes I_U \hspace{1cm}\hbox{and}\hspace{1cm} 
O {\mathcal J} O^T = \rho_E \otimes J_U,
\end{equation}
where $\rho_E = \sum_n c_n |n\rangle\langle n |$ is a density matrix of the environment.  
In the modified-ubit interpretation, our description of Alice's system is given not by $\rho_{UA}$ but rather by the density matrix $\rho_{U'A}$, defined
as
\begin{equation}
\rho_{U'A} = \hbox{Tr}_E \,O \hat{\rho}_\parallel O^T.
\end{equation}
(In this equation we have written $O$ as an abbreviation for $O \otimes I_A$.)
When Eq.~(\ref{newsplit}) is valid, that is, when we can neglect 
contributions to $\nu_j$ of higher than second order, Eq.~(\ref{IJtransform}) implies that 
\begin{equation}
O \hat{\rho}_\parallel O^T = \frac{1}{2}\, \rho_E \otimes \left( I_U \otimes \hbox{Re}\,\sigma + J_U \otimes \hbox{Im}\,\sigma \right)
\end{equation}
and therefore
\begin{equation}
\rho_{U'A} = \frac{1}{2} \left( I_U \otimes \hbox{Re}\,\sigma + J_U \otimes \hbox{Im}\,\sigma \right).
\end{equation}
Note that to this order in perturbation theory, $O\hat{\rho}_\parallel O^T$ is equal to the tensor product $ \rho_E \otimes \rho_{U'A}$ and we can
work out the evolution of $\rho_{U'A}$ without explicitly referring to the environment.  In fact the theory
is almost the same as what we would have gotten simply by applying the complex-to-real prescription of Section II.
The only difference is the constant factor $\xi = 1- \lambda^2/2 $ in the effective Hamiltonian $H' = \xi K_A + iL_A$.  

We now relax the restriction to the second-order expansion of $\nu_j$.  Starting with the same initial state (\ref{initialstate}),
we find that the states Alice can create have the form
\begin{equation}  \label{generalrhohat}
\hat{\rho}_\parallel = \frac{1}{2} \sum_n c_n 
\left( |\Psi_n^+\rangle\langle \Psi_n^+| \otimes \sigma_n+ |\Psi_n^-\rangle\langle \Psi_n^-| \otimes \overline{\sigma}_n \right),
\end{equation}
where each $\sigma_n$ is a positive semi-definite complex matrix with unit trace.  This equation generalizes
Eq.~(\ref{gammaequation}).  Applying our transformation $O$, 
we get
\begin{equation}
O \hat{\rho}_\parallel O^T = \frac{1}{2} \sum_n c_n |n\rangle\langle n| \otimes 
\left( I_U \otimes \hbox{Re}\,\sigma_n + J_U \otimes \hbox{Im}\,\sigma_n \right),
\end{equation}
which is no longer a tensor product between the environment and the $UA$ system.
Now tracing over the environment
gives us
\begin{equation}
\rho_{U'A} = \hbox{Tr}_E\,O\hat{\rho}_\parallel O^T = \frac{1}{2}( I_U \otimes \hbox{Re}\,\sigma + J_U \otimes \hbox{Im}\,\sigma),
\end{equation}
where $\sigma = \sum_n c_n \sigma_n$.  Thus
it remains true that Alice can describe her system by a $d \times d$ complex density matrix $\sigma$.
However, in general this matrix will not evolve according to Eq.~(\ref{newSchrodinger}).  It might decohere, for example.

It is interesting to follow the evolution of $\rho_{U'A}$ numerically for the case of a precessing spin.  
That is, we compute the evolution of $\hat{\rho}_\parallel(t)$ as 
always, but instead of simply tracing over the environment to get $\rho_{UA}$, we apply $O$ before performing the trace, so as to
get $\rho_{U'A}$.  As one would expect from
the results of this section, the ghost effect is no longer seen: over the short term, the Bloch vector associated with $\rho_{U'A}$ retains its full length over 
the whole cycle.  However, the other two effects---the reduced precession frequency and the long-term decoherence---persist and
appear to be the same as before.  In Eq.~(\ref{newSchrodinger}) the reduction in frequency appears as the constant $\xi$ in $H'$.  We do not see the decoherence in this equation because it does not include  
third-order contributions to $\nu_j$.

The notion of a modified ubit raises a new question.  Throughout our analysis we have assumed that
Alice is able to build a device that implements an arbitrary Stueckelbergian $I_E \otimes S_{UA}$ (though as we have seen,
the ubit-environment interaction can render certain components of this operator ineffective).  
However, if Alice can prepare states of the form (\ref{newsplit}), involving
the environment-ubit matrices ${\mathcal I}$ and ${\mathcal J}$, one might wonder whether she also has the ability to
implement a Stueckelbergian of the form $\hat{S} = {\mathcal I}\otimes S_A + {\mathcal J} \otimes T_A$, where $S_A$ is antisymmetric
and $T_A$ is symmetric.  Evidently she has some control over the environment, but does she have the right kind of control 
to be able to implement $\hat{S}$?  It seems likely that the correct answer is no.  Whereas her ability to prepare states of the form (\ref{newsplit})
follows from our initial assumptions, we have not found a way by which she might implement this kind of Stueckelbergian.  
Moreover, the theory is self-consistent as it stands, even though Alice's local Stueckelbergians are 
linear combinations of $I_U$ and $J_U$ in the original basis, while the states she can prepare
can be expressed as such linear combinations
only after we have rotated the $EU$ system by the transformation $O$.  
Indeed, Alice prepares these ``exotic'' states by using 
``ordinary'' Stueckelbergians, relying on the interaction matrix $B_{EU}$ to bring the environment into the picture.

Our modified-ubit interpretation also forces us to extend the argument of Subsection III.E showing that there can be no signaling
through the ubit.  If Alice can prepare states involving the environment, is it possible
that she could leave some mark on the environment that could be read by a second observer?  The answer is no.  
If the second observer does not have access to the $A$ system itself, then that observer's knowledge is contained
in the reduced density matrix $\hbox{Tr}_A\, \hat{\rho}_\parallel$.  For the general form given in Eq.~(\ref{generalrhohat}), this partial
trace is always equal to $(1/2){\mathcal I}$, regardless of what Alice has done.  So Alice cannot send a signal by this 
means.


\section{The Real-Imaginary Split}

In our analysis of the precessing spin, we began with the Hamiltonian $H = \hbar(\Omega/2)Z$ and converted it into
a Stueckelbergian according to the prescription given in Eq.~(\ref{HtoS}), which asks us to split $H$ into its real and imaginary parts.
The result was $S_{UA} = -(\Omega/2)J_U \otimes Z_A$.
But for the very same physical situation, we could just as well have associated the magnetic field direction with the Pauli matrix $Y$ instead of $Z$, in which case
the Hamiltonian would have been $H = \hbar(\Omega/2)Y$ and the Stueckelbergian would have been
$S_{UA} = (\Omega/2)I_U \otimes J_A$.  This latter Stueckelbergian does not involve the ubit matrix $J_U$, and this fact would have made
the predictions of our model different from those we obtained in Sections IV and V.  In particular there would be no decoherence and no retardation 
of the precession.  Moreover
we would get yet another set of results if we chose to assign to the magnetic field direction some other spin matrix such as $(X + Y)/\sqrt{2}$. 
Thus, as we have said before, a given Hamiltonian can give rise to many distinct Stueckelbergians, depending on the basis in 
which the Hamiltonian is written.  

Let us see how this ambiguity is expressed for an $A$ system with dimension $d$.   Given a certain Hamiltonian, we can imagine changing the basis first and then applying Eq.~(\ref{HtoS}).
This procedure would give us, for a specified Hamiltonian $H$, a Stueckelbergian $S_{UA}$ defined by
\begin{equation}  \label{HtoS2}
S_{UA} =  I_U \otimes \, \hbox{Re}\left( - i \, U H U^\dag\right)  + J_U \otimes  \, \hbox{Im}\left( - i \, U H U^\dag\right),
\end{equation}
where $U$ is a unitary transformation representing the change of basis.  (After performing this operation, we could 
perform an orthogonal transformation on the $A$ system, but this second change of basis does not affect the real-imaginary split.)
As the dimension $d$ grows, this freedom 
opens up many new possibilities.  For example, even for a set of non-interacting particles, the complex-to-real correspondence
of Section II could be carried out
in an entangled basis.  And the predicted physics would in general be different for different choices.  

In the specific case of a precessing spin, our model as currently formulated implies that in addition to the magnetic field direction,
an experimenter might in principle also be able to set the orientation of another physically important axis, namely, the axis 
with which we associate the imaginary Pauli matrix $Y$.  The angle between this second axis and the magnetic field direction 
would affect both the rate of precession and the decoherence rate.  

On the other hand, it is conceivable that if something like the ubit model does apply to our world, there could be a law of nature that restricts or determines 
the way in which the real-imaginary split is made, as we suggested in Subsection IV.E.  The simplest possibility would be that the conversion
described by Eq.~(\ref{HtoS}) is required to be carried out 
in a basis in which the Hamiltonian is real.  (This law would have meaning only for an isolated system.)  
Then the Stueckelbergian would be uniquely determined by the
Hamiltonian---up to an orthogonal transformation on the $A$ system---and it would always take
the form
\begin{equation}
S_{UA} = -J_U \otimes K_A.
\end{equation}
In this case the ubit would be playing a role very much like that of the phase factor $e^{-iEt/\hbar}$ for a state with definite
energy $E$.  In the example of a 
precessing spin, in our alternative interpretation there would be no need to specify a second axis as in the preceding paragraph.  
The reduced precession frequency and the decoherence would always be just as we 
described them in Sections IV and V, because the direction of the magnetic field would always be associated with a real spin matrix.
Moreover, for an $A$ system of arbitrary dimension, the effective Hamiltonian of Section VI would always have the form
$H' = \xi K_A$, where $K_A$ is a symmetric real matrix---there would be no imaginary part $iL_A$.  One consequence is that
{\em all} processes would be slowed by exactly the same factor $\xi$, in which case the slowing would be
unobservable.  Thus if we (i) adopt the modified-ubit interpretation of Section VI and (ii) assume that Eq.~(\ref{HtoS}) is always
to be applied in a basis in which the Hamiltonian is real, then of the deviations from standard quantum theory we have
identified, the only one that could actually be observed is the decoherence.

\section{Discussion}

Over the years a number of researchers have taken an interest in real-vector-space quantum theory.  In a paper published
around the same time as Stueckelberg's papers on the subject, Dyson argued that when we make use of the time-reversal
operator, we are implicitly basing our theory on the field of real numbers, since that operator is antilinear
in the complex theory but can be expressed as a linear operator in the real theory \cite{Dyson}.  (See also Ref.~\cite{Baez}.)
More recently Gibbons and others have argued that the complex structure might be an emergent property associated with
the emergence of a time direction \cite{Gibbons, Gibbons2}.  Myrheim has observed that a real-vector-space theory could allow a canonical commutation relation of
the form $[x,p] = J\hbar$ even for a finite-dimensional state space \cite{Myrheim}.  
There has also been work on real vector spaces in quantum information theory \cite{Rudolph, Fernandez, McKague1, McKague2, Rungta, Vlasov, HardyWootters, Wootters}, including the proof mentioned earlier
that $n+1$ rebits can simulate $n$ qubits.  

In this paper we have considered a specific model within real-vector-space quantum theory, characterized by
the inclusion of 
a single binary object, the ubit $U$, which is not localized and which can interact with anything in the universe.  We have 
focused on characterizing the effective theory describing the behavior of a system consisting of the ubit and
a local real-vector-space object $A$.  
From the outset we have assumed that our local observer Alice has the ability to prepare any initial state of the $UA$ system
and could build an apparatus that would implement any Stueckelbergian $S_{UA}$.  However, if both ${s}$---the size
of a typical eigenvalue of ubit-environment interaction---and the ubit's rotation rate $\omega$ are in effect infinitely large compared to any
local frequency, then the local observer's abilities are severely constrained.
Assuming a large environment, the interaction between the ubit and the environment has the effect of enforcing Stueckelberg's rule
on the operators of the $UA$ system.  If Alice
tries to prepare a state $\rho_{UA}$ that does not commute with $J_U \otimes I_A$, the noncommuting part will instantly
disappear.  (The noncommuting part has zero trace---so its disappearance does not entail a loss of probability.)  And if she tries to use a Stueckelbergian $S_{UA}$ that does not commute with 
$J_U \otimes I_A$, its noncommuting part will have no observable effect.  Note that in our model neither the state $\hat{\rho}$ of the whole 
universe nor the full Stueckelbergian $\hat{S}$ commutes with $I_E \otimes J_U \otimes I_A$.  So from a global perspective the theory 
is quite different from standard quantum theory.  It is only at the local level that we see all operators commuting with $J_U$.  

Moreover, this automatic enforcement of Stueckelberg's rule at the local level does not cause the effective theory to be equivalent to standard quantum mechanics,
because the ubit-environment interaction has other effects as well.
As we have seen, the degree of divergence between the ubit model and standard quantum theory depends on how
one interprets the ubit model.  We have distinguished two interpretations, which we have called the literal interpretation
and the modified-ubit interpretation.  

The literal interpretation assumes that $\rho_{UA}$ is the full and correct description of the state observed by Alice.  
If we adopt this interpretation, then
in addition to losing the states and transformations that would have violated Stueckelberg's rule, we also 
lose certain states and transformations that quantum mechanics normally allows.  For example, for the spin
of a spin-1/2 particle, the Bloch sphere is in effect flattened into an oblate ellipsoid.  Except in a certain preferred plane,
it is impossible to prepare a pure state of spin in any direction.  In particular, in standard quantum theory there will always be one
direction with which we associate a purely imaginary spin matrix (usually called the $y$ direction).  This is the direction in which,
according to the literal interpretation, the purity of a spin state is the most limited.  Moreover, if one prepares a pure state in the 
favored plane and allows the spin to precess out of that plane, then the purity of the state will temporarily be diminished in order
for the Bloch vector to be able to fit inside the flattened sphere.  

One can see how the ubit model thus provides a kind of interpolation between standard complex-vector-space
quantum theory and standard real-vector-space quantum theory with no ubit.  In the former case, we have the full Bloch sphere. 
In the latter case, we have a ``Bloch disk,'' as all rebit states are confined to a plane.  In the ubit model, in the literal interpretation,
the width of the narrow axis of the flattened Bloch sphere depends on the parameter $\lambda = {s}/\omega$.  As this parameter
gets larger---e.g., as the ubit's rotation rate diminishes---the Bloch ellipsoid becomes more flattened.  Indeed, when the ubit is not
rotating at all, there is nothing preventing the ubit's interaction with the environment from always bringing $\rho_{UA}$ to the
form $(1/2)I_U \otimes \rho_A$, in which case the ubit becomes inconsequential and, since $\rho_A$ is real, the spin is confined to the Bloch
disk.  

However, there are good reasons to prefer the modified-ubit interpretation.  In this interpretation the matrix $\rho_{UA}$ does not describe
what Alice sees.  Instead, the state of Alice's system is described by
$\rho_{U'A} = \hbox{Tr}\,O\hat{\rho}_\parallel O^T$, where $O$ is an orthogonal transformation on the 
$EU$ system.  In the case of a spin-1/2 particle, in the 
alternative interpretation the Bloch sphere is not flattened: the set of possible density matrices $\rho_{U'A}$
includes all the pure states on the sphere.  In effect the density matrix $\rho_{U'A}$ includes 
the ghost part that would have disappeared upon simply
tracing $\hat{\rho}_\parallel$ over the environment to get $\rho_{UA}$.  
It makes sense to describe Alice's states in terms of $\rho_{U'A}$, because this 
description reflects what Alice is actually able to do---what states she can prepare and test for---according to our original assumptions.  
We have noted that though Alice's actions involve the environment, they do so in a way that does not convey information.  

The alternative interpretation yields a greater degree of 
agreement between the ubit model and standard quantum theory.  To second order
in our parameter $\lambda = {s}/\omega$, the only difference between the two theories, with regard to 
the behavior of the local system, is a factor $\xi \lsim 1$ multiplying the real part of the Hamiltonian.  
Moreover, we have also contemplated the possibility of a law of nature that would force the local Stueckelbergian to be of the form
$-J_U \otimes K_A$, in which case the factor $\xi$ becomes a universal retardation factor and 
therefore unobservable.

As we go to higher order in $\lambda$ the correspondence
breaks down further, and the ubit model will surely be very different from standard quantum theory once $\lambda$ is of order
unity.  
One difference we have been able to identify is the spontaneous decoherence, which appears as a third-order effect.  
For the case of a precessing qubit, we have an analytic estimate of the coherence time in our model: it is 
$\tau \approx 1/(\lambda^3 \Omega)$, where $\Omega$ is the angular frequency of precession.  
Thus experimental observations of the product $\tau\Omega$ can be used to place an upper bound on 
$\lambda$.  A particularly high value of $\tau\Omega$ was obtained
recently by Chou {\em et al.}~for an optical atomic transition \cite{Chou}.  These authors observed a coherence time around 10 seconds for a 
transition with angular frequency $\Omega = 7 \times 10^{15}\,$s$^{-1}$.  This result implies that our parameter $\lambda$
cannot be much larger than $10^{-6}$.  Note that the slowing by the factor $\xi$, being a second-order effect, could conceivably
place a tighter bound on $\lambda$, but because of our uncertainty about the real-imaginary split, it is hard to 
know how to look for this slowing.  

Again, the model we have described here is highly nonlocal, and it is interesting to ask whether we would get 
similar results from a local theory in which the ubit is replaced with a ubit field.  (One is reminded
of Refs.~\cite{McKague1, McKague2}, in which the authors show that a complex-vector-space quantum computation
can be simulated {\em locally} by a real-vector-space quantum computation.)  We have relied on the environment to 
effectively cut out the parts of the local Stueckelbergian and local density matrix that do not satisfy Stueckelberg's rule.  In order
to get this kind of environmental effect with a ubit field, we would probably need the field to be such that a change at one location
would quickly 
bring a large part of the
surrounding
environment into the dynamics to create effects similar to what we have seen here.  

Clearly the work presented here leaves many questions unanswered.  It is only an initial exploration of the model.
But we can at least see that the model is not in obvious conflict with observation as long as $\lambda$ is sufficiently
small.  Note that in a complete theory it should not be necessary that this parameter remain constant for all time.  
So the degree of deviation from standard quantum theory could conceivably be different at different stages of the universe's evolution.  
The effective theory might be experimentally indistinguishable from standard quantum theory at the present time but
could look much more like ordinary real-vector-space quantum theory (with no ubit) at some very early time.  
In this sense the familiar complex structure of quantum theory could indeed be an emergent feature in the ubit model.  

\bigskip
\bigskip
\bigskip

\noindent {\bf ACKNOWLEDGMENTS}

\bigskip
\medskip

The research presented here and the paper itself have both benefitted significantly from comments and 
suggestions by Fred Strauch and Dave Tucker-Smith.  We also thank
the members of the Quantum Foundations Group at Perimeter Institute, and members of the 
Institute of Mathematical Sciences in Chennai, for their valuable comments and questions.

\bigskip
\bigskip
\bigskip

\noindent {\bf APPENDIX A: BASIC PERTURBATION EXPANSION}

\bigskip
\medskip

To summarize our notation: the Hermitian operator on which we base our perturbation expansions is
\begin{equation}  \label{Gagain}
G = G_0 +\lambda  V,
\end{equation}
where $\lambda = s/\omega$ and the unperturbed matrix
$G_0$ has only two distinct eigenvalues, $\pm 1$, each corresponding to an $N$-dimensional subspace.
The states $|\Phi^\pm_n\rangle$ are the eigenstates of $G_0$ that diagonalize $V$ within each of these two
subspaces, and $|\Psi^\pm_n\rangle$ are the corresponding eigenstates of $G$.  The exact eigenvalues of $G$ are labeled $\pm g_n$.  

To do our perturbation calculations, we need to expand $g_n$ out to third order and 
$|\Psi^\pm_n\rangle$ out to second order.  We write here the expansions for $g_n$ and $|\Psi^+_n\rangle$.  As we have 
mentioned earlier, the eigenvector $|\Psi^-_n\rangle$ is the complex conjugate of $|\Psi^+_n\rangle$.  
Our expansion for the eigenvectors is derived from Eqs.~(141--143) of Ref.~\cite{Hirschfelder}.  
Our expansion for the eigenvalues is obtained from the expanded eigenvectors together with Eq.~(7) of the same paper.  
The normalization convention in that paper is such that the expanded vectors are not of unit length, but in the expansion 
of $|\Psi_n^+\rangle$ below we
have renormalized the vector to unit length (to second order in $\lambda$), which introduces the two second-order terms proportional
to $|\Phi_n^+\rangle$.    Here we use the notation $M_{nk}$
as an abbreviation for $\langle \Phi_n^+ |M|\Phi_k^+\rangle$, for any matrix $M$.  Again, $P_-$ is the projection onto the space
spanned by the states $|\Phi^-_n\rangle$.  

The expansion for $g_n$ is
\begin{equation}
\begin{split}
g_n &= 1 + \lambda V_{nn} + \frac{\lambda^2}{2}(VP_-V)_{nn} \\
& + \frac{\lambda^3}{4}\left[ (VP_-VP_-V)_{nn} - V_{nn}(VP_-V)_{nn} +  {\sum_k}^{'} \frac{\left| (VP_-V)_{nk} \right|^2}{V_{nn} - V_{kk}}  \right],
\end{split}
\end{equation}
where the prime on the summation sign indicates that $k$ runs over all values other than $n$.  

The expansion for $|\Psi_n^+\rangle$ is
\begin{equation}
\begin{split}
|\Psi_n^+\rangle &= |\Phi_n^+\rangle + 
\frac{\lambda}{2}\left[ P_-V|\Phi_n^+\rangle + {\sum_k}^{'} \frac{(VP_-V)_{kn}}{V_{nn} - V_{kk}}|\Phi_k^+\rangle \right] \\
& + \frac{\lambda^2}{4} \Bigg[ P_-VP_-V|\Phi_n^+\rangle - V_{nn} P_-V |\Phi_n^+\rangle - \frac{1}{2}(VP_-V)_{nn}|\Phi_n^+\rangle \\
& \hspace{10mm} + {\sum_k}^{'} \bigg( \frac{(VP_-V)_{kn}}{V_{nn}-V_{kk}}P_-V |\Phi_k^+\rangle
- \frac{(VP_-V)_{kn}(VP_-V)_{nn}}{(V_{nn} - V_{kk})^2} |\Phi_k^+\rangle \\
& \hspace{23mm}-\frac{1}{2} \frac{ \left| (VP_-V)_{kn} \right|^2}{(V_{nn} - V_{kk})^2} |\Phi_n^+\rangle  
 + \frac{(VP_-VP_-V)_{kn}}{V_{nn} - V_{kk}} |\Phi_k^+ \rangle \\
& \hspace{23mm} - \frac{(VP_-V)_{kn}V_{nn}}{V_{nn} - V_{kk}} |\Phi_k^+\rangle 
+ {\sum_l}^{'} \frac{ (VP_-V)_{kl} (VP_-V)_{ln} }{ (V_{nn} - V_{kk}) (V_{nn} - V_{ll}) } |\Phi_k^+\rangle \bigg) \Bigg] . 
\end{split}
\end{equation}

\bigskip
\bigskip
\bigskip

\noindent {\bf APPENDIX B: LETTING $\omega$ APPROACH INFINITY}

\bigskip
\medskip

For most of this paper we have assumed that both $s$ and $\omega$ are very large compared to $S_{UA}$.  We have also assumed
that the environment dimension $N$ becomes arbitrarily large.  In this Appendix we consider
a different limit.  Here $s$ and $N$ both remain finite, and we let $\omega$ go to infinity.  Our aim is to show that in this limit,
the evolution operator in the ubit model, $e^{\hat{S}t}$, becomes equivalent to a corresponding evolution operator
of standard quantum theory.  

Again, our Stueckelbergian for the whole $EUA$ system is 
\begin{equation}
\hat{S} = - \omega I_E \otimes J_U \otimes I_A + s B_{EU}\otimes I_A + I_E \otimes S_{UA}.
\end{equation}
Let us now write this operator as
\begin{equation}
\hat{S} = - \omega \hat{J} + \hat{D},
\end{equation}
where $\hat{J} = I_E \otimes J_U \otimes I_A$ and $\hat{D} = s B_{EU}\otimes I_A + I_E \otimes S_{UA}$.  

We begin by rewriting $e^{\hat{S}t}$ as follows, using the fact that $\hat{S}$ is antisymmetric so that $\hat{S}^T\hat{S} = -\hat{S}^2$:
\begin{equation}  \label{bigexpansion}
\begin{split}
e^{\hat{S}t} &= I + \hat{S}t + \frac{1}{2!}\hat{S}^2 t^2 + \frac{1}{3!} \hat{S}^3 t^3 + \cdots \\
&=\left( I - \frac{1}{2!}\hat{S}^T\hat{S}\,t^2 + \cdots \right) + \left( \hat{S}t - \frac{1}{3!} \hat{S}\,\hat{S}^T\hat{S}\,t^3 + \cdots \right) \\
&=\hat{S}\,\frac{\sin\big(\sqrt{\hat{S}^T\hat{S}} \, t \big)}{\sqrt{\hat{S}^T\hat{S}}} + \cos\big( \sqrt{\hat{S}^T\hat{S}}\, t \big).
\end{split}
\end{equation}
Note that $\sqrt{\hat{S}^T \hat{S}}$ is well defined since $\hat{S}^T\hat{S}$ is a positive semi-definite matrix.  Now we make the substitution
$\hat{S} = -\omega \hat{J} + \hat{D}$:
\begin{equation}
\hat{S}^T\hat{S} = \big( -\omega \hat{J}^T + \hat{D}^T \big) \big( -\omega \hat{J} + \hat{D}\big) = \omega^2 I +
 \omega\big(\hat{J}\hat{D} + \hat{D}\hat{J} \big) - \hat{D}^2.
\end{equation}
Here we have used the fact that both $\hat{J}$ and $\hat{D}$ are antisymmetric.  Recall that the part of $\hat{D}$ that commutes
with $\hat{J}$ can be written as $\hat{D}_c = (1/2)\big(\hat{D} - \hat{J}\hat{D}\hat{J}\big)$.
In terms of $\hat{D}_c$, we have
\begin{equation}
\hat{S}^T \hat{S} = \omega^2 I + 2 \omega \hat{J}\hat{D}_c - \hat{D}^2,
\end{equation}
so that
\begin{equation}
\sqrt{\hat{S}^T\hat{S}} = \omega \sqrt{ I + \frac{2 \hat{J}\hat{D}_c}{\omega} - \frac{ \hat{D}^2 }{\omega^2} }.
\end{equation}
We can therefore write the evolution operator (Eq.~(\ref{bigexpansion})) as 
\begin{equation}  \label{hairyevolution}
e^{\hat{S}t} = \bigg(- \hat{J} + \frac{\hat{D}}{\omega} \bigg)
 \frac{ \sin\Big( \omega t \sqrt{ I + \frac{2 \hat{J} \hat{D}_c }{\omega}  - \frac{\hat{D}^2}{\omega^2} } \Big) }{\sqrt{ I + \frac{2 \hat{J} \hat{D}_c }{\omega}  - \frac{\hat{D}^2}{\omega^2} } } 
 + \cos \left(  \omega t \sqrt{ I + \frac{2 \hat{J} \hat{D}_c }{\omega}  - \frac{\hat{D}^2}{\omega^2} } \right).
\end{equation}

When $\omega$ is very large, we can expand the square roots in powers of $1/\omega$.  Ignoring terms of order $1/\omega$ 
multiplying the sine, and also ignoring terms of order $1/\omega$ inside the sine and cosine, 
we get
\begin{equation}
e^{\hat{S}t} \approx -\hat{J} \sin\big[ \big( \omega I + \hat{J}\hat{D}_c  \big) t \big] + \cos \big[ \big( \omega I + \hat{J}\hat{D}_c  \big) t \big] 
= e^{ ( - \hat{J} \omega + \hat{D}_c ) t } = e^{- \hat{J} \omega t} e^{\hat{D}_c t}.
\end{equation}
Thus, when $\omega$ is very large, the dynamics effectively separates into two parts: (i) the ubit rotates very rapidly, and (ii)
the whole system evolves according to the Stueckelbergian $\hat{D}_c$.  (These two operations commute with each other.)  

We now want to show that this evolution amounts to an ordinary quantum mechanical evolution.
The Stueckelbergian $\hat{D}_c$ is exactly what
one would get by starting with the Hamiltonian 
\begin{equation}
H =  \big(\hat{D}_c\big)_{00} + i\big( \hat{D}_c \big)_{10}
\end{equation} 
and simply rewriting the same physics in real-vector-space terms as in Section II.  Here the indices 0 and 1 are ubit indices as in that section.  When 
we write out the definition of $\hat{D}$, this Hamiltonian becomes
\begin{equation}
H = H_{\mathcal E}\otimes I_{\mathcal A} + I_{\mathcal E} \otimes H_{\mathcal A},
\end{equation}
where $H_{\mathcal E} = s\big\{ \big[(B_{EU})_c\big]_{00} + i \big[(B_{EU})_c \big]_{10}\big\}$
and $H_{\mathcal A} =  \big[(S_{UA})_c\big]_{00} + i \big[(S_{UA})_c \big]_{10}$, and we are using script letters to refer to
systems described in terms of a complex vector space.
(Again the subscript $c$ means that we are taking only the part of the operator that commutes with $\hat{J}$.)  
Thus the environment and the local system evolve independently, each according to its own Hamiltonian.  

We conclude that, even without letting $s$ or $N$ go to infinity, the effective dynamics of the local system in the ubit model reduces to the dynamics 
of standard quantum mechanics as $\omega$ approaches infinity.  On the other hand, if we keep terms of order $1/\omega$ 
in Eq.~(\ref{hairyevolution}), we obtain correction terms that do not commute with $\hat{J}$ (for a generic $\hat{D}$).  So we do not expect
a perfect correspondence with standard quantum theory for any finite value of $\omega$.

\bigskip
\bigskip
\bigskip

\noindent {\bf APPENDIX C: CONSTRUCTING THE ORTHOGONAL TRANSFORMATION $O$}

\bigskip
\medskip

In this Appendix our aim is to find an orthogonal transformation $O$ on the $EU$ system such that
$O|\Psi_n^+\rangle\langle\Psi_n^+| O^T = (1/2) | n\rangle\langle n| \otimes (I_U -i J_U)$, where
the real vectors $|n\rangle$ constitute an orthonormal basis for the environment.  We construct
$O$ in two steps.  First, let $O_1$ be the matrix
\begin{equation}
O_1 = \sum_n \left(|\Phi_n^+\rangle\langle\Psi_n^+| + |\Phi_n^-\rangle\langle\Psi_n^-|\right).
\end{equation}
$O_1$ transforms between two orthonormal bases, so it is unitary.  It is also real and therefore orthogonal. 
Upon applying $O_1$ to $|\Psi_n^+\rangle\langle\Psi_n^+|$, we get
\begin{equation}  \label{O1effect}
O_1 |\Psi_n^+\rangle\langle\Psi_n^+| O_1^T = |\Phi_n^+\rangle\langle \Phi_n^+| = |\phi_n^+\rangle\langle\phi_n^+| \otimes
|+\rangle\langle +| = \frac{1}{2}\,|\phi_n^+\rangle\langle\phi_n^+| \otimes (I_U - iJ_U),
\end{equation}
where we are using the factorization $|\Phi_n^+\rangle = |\phi_n^+\rangle \otimes |+\rangle$ introduced in
Subsection III.C.  

Next, let $U$ be the unitary matrix that takes $|\phi_n^+\rangle$ to $|n\rangle$.  That is, $U = \sum_n |n\rangle\langle \phi_n^+ |$.  
And let $O_2$ be the real-vector-space version of $U$ according to the transcription rules of Section II.  That is, 
\begin{equation}
O_2 = \hbox{Re}\,U \otimes I_U + \hbox{Im}\,U \otimes J_U.
\end{equation}
By rewriting $U |\phi_n^+\rangle\langle\phi_n^+| U^\dag = |n\rangle\langle n|$ in real-vector-space terms, we get
\begin{equation}  \label{C1}
O_2 \left( \hbox{Re}\,|\phi_n^+\rangle\langle\phi_n^+| \otimes I_U + \hbox{Im}\,|\phi_n^+\rangle\langle\phi_n^+| \otimes J_U \right) O_2^T
= |n\rangle\langle n| \otimes I_U.
\end{equation}
Multiplying both sides of this equation by $-i I_E \otimes J_U$ (which commutes with $O_2$) gives us
\begin{equation}  \label{C2}
O_2 \left( i\, \hbox{Im}\,|\phi_n^+\rangle\langle\phi_n^+| \otimes I_U -i\,  \hbox{Re}\,|\phi_n^+\rangle\langle\phi_n^+| \otimes J_U \right) O_2^T
= -i\, |n\rangle\langle n| \otimes J_U.
\end{equation}
Now we add Eqs.~(\ref{C1}) and (\ref{C2}) to get
\begin{equation}  \label{O2effect}
O_2 \left[ |\phi_n^+\rangle\langle\phi_n^+| \otimes (I_U -i J_U ) \right] O_2 = |n\rangle\langle n | \otimes (I_U -i J_U),
\end{equation}
in which the left-hand side mirrors the right-hand side of Eq.~(\ref{O1effect}).  Finally we define $O$ to be 
$O = O_2O_1$.  Then Eqs.~(\ref{O1effect}) and (\ref{O2effect}) imply that
\begin{equation}
O |\Psi_n^+\rangle\langle\Psi_n^+ | O^T = \frac{1}{2}\, |n\rangle\langle n | \otimes (I_U - i J_U),
\end{equation}
which is what we wanted to show.  

To get some mathematical insight it is interesting to work out the effect of $O$ on $S_{EU}$, though in our modified-ubit interpretation we
do not perform this transformation.  We can write $S_{EU}$ as
\begin{equation}
S_{EU} = - i\omega \sum_n g_n \left( |\Psi_n^+\rangle\langle \Psi_n^+ | - |\Psi_n^-\rangle\langle \Psi_n^- | \right),
\end{equation}
where the $g_n$'s are eigenvalues of $G$ as in Subsection III.B.  Upon applying $O$, we get
\begin{equation}
O S_{EU} O^T = - \omega \Big( \sum_n g_n |n\rangle\langle n| \Big) \otimes J_U.
\end{equation}
Thus the transformation $O$ brings $S_{EU}$ to a tensor-product form.  Any antisymmetric real matrix such as
$S_{EU}$ can be brought to block-diagonal form by an orthogonal transformation, with $2\times 2$ blocks 
proportional to $J$ \cite{Horn}.  If we 
take the basis defined by the $|n\rangle$'s to be the standard basis for the environment, then $O$ is 
an orthogonal transformation that brings $S_{EU}$ to this form.

\end{document}